%% file: main.tex
\documentclass[onefignum,onetabnum]{siamart171218}
\usepackage{float}
\usepackage{placeins}
\usepackage{booktabs}  % Add this in your preamble for \midrule, \cmidrule, etc.
\usepackage{makecell}
\input{ex_shared}
\ifpdf
\hypersetup{
  pdftitle={Model Error {Covariance} Estimation for Weak Constraint Data Assimilation},
  pdfauthor={S. R. Babyale, J. Mead, D. Calhoun and P. Azike}
}
\fi

\usepackage{xcolor}
\definecolor{RED}{named}{red}
\usepackage{hyperref}

\makeatletter
\let\NR@oldlabel\label
\makeatother
\begin{document}

\maketitle

% REQUIRED
\begin{abstract}
  State estimates from weak constraint four-dimensional variational (4D-Var) data assimilation can vary significantly depending on the data and model error covariances. As a result, the accuracy of these estimates heavily depends on the correct specification of both model and observational data error covariances. In this work, we assume that the data error is known and and focus on estimating the model error covariance by framing weak constraint 4D-Var as a regularized inverse problem, where the inverse model error covariance serves as the regularization matrix.We consider both isotropic and non-isotropic forms of the model error covariance, with hyperparameters such as the model error variance, spatial correlation length, and temporal correlation scale. Using the representer method, we reduce the 4D-Var problem from state space to data space, enabling the efficient application of regularization parameter selection  techniques to estimate the model covariance hyperparameters. The Representer method also provides an analytic expression for the optimal state estimate, allowing us to derive matrix expressions for the three regularization parameter selection methods i.e. the L-curve, generalized cross-validation (GCV), and the $\chi^2$ method. We validate our approach by assimilating simulated data into a 1D transport equation modeling wildfire smoke transport under various observational noise and forward model perturbations. In these experiments the goal is to identify the model error covariance estimates that accurately capture the influence of observational data versus model predictions on assimilated state estimates. The regularization parameter selection methods successfully estimate hyperparameters for both isotropic and non-isotropic model error covariances, that reflect whether the first guess model predictions are more or less reliable than the observational data. The results further indicate that isotropic variances are sufficient when the first guess is more accurate than the data whereas non-isotropic covariances are preferred when the observational data is more reliable.

\end{abstract}

% REQUIRED

% REQUIRED

\section{Introduction} \label{sec:Int}

Data assimilation is an important tool for analyzing complex physical phenomena across various scientific disciplines, such as geosciences, oceanography, and atmospheric sciences, among others \cite{carrassi2018data}. At its core, data assimilation merges theory, mathematical models, and observations to provide a more accurate estimation of a system's state. While dynamical models provide insights into physical interactions and observations provide point measurements, combining both through data assimilation results in a more robust and realistic depiction of the physical phenomena in play. This integration offers a clear advantage over relying solely on estimates from either the model or observations alone as models often face numerous limitations and challenges and observations are frequently sparse and incomplete \cite{bannister2017review,benedetti2009aerosol, hyer2023data,  kalnay2003atmospheric, lorenc2000met, otkin2021nonlinear}.

Data assimilation methods are broadly categorized into sequential and variational approaches. Sequential methods, such as the Particle Filter, Kalman Filter, Extended Kalman Filter (EKF), and Ensemble Kalman Filter (EnKF), are based on stochastic filtering principles and Bayesian minimum variance estimation. These methods sequentially update the system state statistics as new observations become available, propagating the state forward in time \cite{evensen1994sequential, gordon1993novel, papadakis2010data, tandeo2020review}. Among these, the Kalman filter, assumes that model and observational errors follow Gaussian distributions and propagates error statistics dynamically with each update \cite{evensen2009rank, kalman1960new}. While sequential methods are computationally efficient and widely used in real-time applications, they face challenges in dealing with long time windows and strongly nonlinear systems \cite{anderson1999monte}.

Variational methods, on the other hand, are rooted in optimal control theory and calculus of variations. Examples include three-dimensional variational data assimilation (3D-Var) and four-dimensional variational data assimilation(4D-Var). The 3D-Var approach estimates the state of a given system at a single time step by minimizing a cost function that balances the deviation of the state from the background state as well as the misfit between the state and the observations. The 4D-Var approach extends 3D-Var to include the temporal dimension, allowing for the assimilation of observations over a given time window, making it particularly suitable for systems with complex temporal dynamics  \cite{bennett1992inverse, courtier1994strategy, le1986variational, rabier1992four}. Unlike the Kalman filter, which continuously updates the state as new observations arrive, 4D-Var computes the optimal trajectory across the entire assimilation window.

In 4D-Var data assimilation, the model is corrected in both spatial and temporal dimensions using either strong or weak constraint formulations. The strong constraint formulation assumes that there are no errors in the model dynamics, attributing all discrepancies to observational errors. Conversely, the weak constraint formulation explicitly acknowledges model errors, which may arise from parameterization, unresolved processes, uncertainties in the forcing, etc \cite{vidard2004variational}. Weak constraint 4D-Var allows for corrections in both spatial and temporal dimensions by incorporating model error terms into the assimilation process, leading to improved estimates, especially in large-scale geophysical models where model errors are significant \cite{fisher2011weak, tr2006accounting}. Although, model error was often neglected in variational data assimilation under the assumption that its influence was minor compared to errors in initial condition and observations, recent studies have shown that incorporating model error significantly improves assimilation accuracy \cite{tremolet2007model, zupanski1997general}. 

The accuracy of data assimilation state estimates strongly relies on specifying appropriate observational and model (or background) error covariances. These error covariances determine the magnitude and structure of corrections applied to their respective trajectories, thereby specifying the balance between the model and observational information. A comprehensive review of methods that estimate error covariances in ensemble data assimilation is given \cite{tandeo2020review}. Here we introduce a global approach to estimating model error covariance over the entire assimilation time range. We assume that the observation error covariance is well-characterized and focus on estimating model error covariance which is more challenging to specify due to the high dimensionality of the systems and the complexity of model dynamics and external forcings.

In this work, we propose a principled approach for estimating model error covariance by treating weak constraint 4D-Var as a regularized inverse problem, where the inverse model error covariance acts as a regularization matrix. Inverse problems, of which data assimilation is fundamentally a subset, aim to estimate unknown parameters or states from observable data. While inverse methods typically invert the forward model for unknown parameters, data assimilation typically focuses on estimating system states from measured outputs of complex systems. Inverse problems can be classified as either well-posed or ill-posed. A problem is well-posed if a solution exists, is unique, and changes continuously with the input data, whereas a problem is ill-posed if it lacks one or more of these properties. In practice, most inverse problems are ill-posed. The most common cause of ill-posedness is non-uniqueness, which typically arises when the number of parameters (or states) to be estimated exceeds the amount of available observational data, which is often the case in data assimilation problems. To address this, regularization techniques are used to incorporate additional information or constraints, thereby stabilizing the solution \cite{evensen1994inverse, sanz2018inverse, zhou2014inverse}. 

Common regularization techniques include Bayesian methods, which incorporate prior knowledge about the parameters into the solution process, and Tikhonov regularization, which adds a penalty term to control the magnitude of the parameters. Both techniques are often formulated as a least-squares problems with an $L^2$-norm regularization term. In our setting, regularization of 4D-Var is imposed through the inverse model error covariance, and the strength and structure of this regularization depend on hyperparameters such as model error variance, spatial correlation length, and temporal correlation scale.

Regularization parameter selection methods are used to determine the influence of the regularization term and balance the trade-off between it and the fit to the observational data. Methods such as the L-curve, Generalized Cross-Validation (GCV), and the $\chi ^2$ method provide systematic ways to identify this balance. The L-curve method involves plotting the norm of the solution against the norm of the residuals to identify a corner point that represents their optimal balance \cite{hansen1999curve}. Generalized Cross-Validation (GCV) minimizes the prediction error by systematically leaving out parts of the data and estimating the error \cite{golub1979generalized}. The $\chi ^2$ method ensures that the regularized residual behaves like the expected distribution \cite{mead2008parameter, Mead_Hammerquist_2013, Mead_2020,renaut2010regularization}. These methods ensure that the chosen regularization parameter effectively stabilizes the solution while maintaining an accurate fit to the observational data, thereby enhancing the overall solution's stability and accuracy \cite{aster2018parameter, engl1994using, golub1999tikhonov,     tikhonov1977solutions}. 

We leverage the regularization parameter selection methods(L-curve, Generalized Cross-Validation (GCV), and the $\chi ^2$ method) to estimate dynamic model error covariance hyperparameters. The elegance of the representer method seamlessly integrates with these regularization parameter selection techniques, enabling efficient computations. By estimating the model error covariance in this way, our approach paves the way for robust and reliable state estimation through weak-constraint 4D-Var data assimilation, even in the face of substantial model and observational uncertainties.

This article is structured as follows: In \cref{sec:2},we formulate variational data assimilation as a regularized inverse problem and apply regularization parameter selection methods to estimate hyperparameters for the background and model error covariances in 3D-Var and 4D-Var respectively. In \cref{sec:3}, we present an overview of weak constraint 4D-Var using the method of representers and establish regularization parameter selection methods for estimating hyperparameters for both isotropic and non-isotropic model error covariances. Numerical experiments demonstrating the effectiveness of this approach are presented in \cref{sec:4}. Finally, we conclude in \cref{sec:6}.

%%%%%%%%%%%%%%%%%%%%%%%%%%%%%%%%%%%%%%%%%%%%%%%%%%%%%%%%%%%%%%%%%%%%%%%%%%%%%%%%%%%%%%%%%%%%%%%%%%%%%%%%%%%%%%%%%%%%%%%%%%%%%%%%%%%%%%%%%%%%%%%%%%%%%%%%%%%%%%

\section{Variational Data Assimilation as Regularization}\label{sec:2}

We frame variational data assimilation as Tikhonov regularization. This allows us to treat background and model error covariance estimation as regularization parameter selection problems.  

\subsection{3D-Var background error variance estimation}\label{subsec_2.1}

In the static case of variational assimilation, the 3D-Var method estimates the system state $\mathbf{x} \in \mathbb{R} ^N$ at a single time by minimizing the cost function:
\begin{equation}
 \mathcal J(\mathbf{x}) = \frac{1}{2} (\mathbf{x} - \mathbf{x}_b)^T \mathbf{B}^{-1} (\mathbf{x} - \mathbf{x}_b) + \frac{1}{2} (\mathbf{d} - \mathbf{H}\mathbf{x})^T \mathbf{R}^{-1} (\mathbf{d} - \mathbf{H}\mathbf{x}) \label{A1},
\end{equation}
where $\mathbf{x}_b \in \mathbb{R} ^N$ is the background state, $\mathbf{B}$ is the background error covariance matrix, $\mathbf{d} \in \mathbb{R} ^ M$ represents the observational data, $\mathbf{R}$ denotes the observation error covariance matrix and $\mathbf{H}$ is the observation operator that maps the state space to the observation space. This formulation can be viewed as Tikhonov regularization, where $\mathbf{x} - \mathbf{x}_b$ serves as the regularization term and $\mathbf B^{-1}$ as the regularization matrix. Minimizing the cost function gives the optimal state estimate $\hat {\mathbf{x}}$ as:
\begin{equation}
\hat {\mathbf{x}} = \mathbf{x}_b + \mathbf{B}\mathbf{H}^T(\mathbf{H}\mathbf{B}\mathbf{H}^T+\mathbf{R})^{-1} (\mathbf{d} - \mathbf{H}\mathbf{x}_b) \label{2.2}.
\end{equation}

The accuracy of the state estimate $\hat{\mathbf{x}}$ depends on proper specification of the background and observation error covariances, $\mathbf{B}$ and $\mathbf{R}$. While $\mathbf{R}$ can often be derived from the characteristics of the data, such as instrument errors or inferred using standard error models, specifying $\mathbf{B}$ is more challenging due to model uncertainties and limited prior knowledge. If incorrectly estimated, an overly large $\mathbf{B}$ may cause the solution to overfit noisy data or the problem remains ill-posed, leading to instability and unrealistic results. Conversely, an excessively small $\mathbf{B}$ can overly smooth the solution, potentially ignoring critical features in the data.

To address this, we use regularization parameter selection techniques to estimate the background error covariance $\mathbf{B}$. Assuming that background errors are uncorrelated and have equal variance in all directions (isotropic), i.e., $\mathbf{B} = \sigma_b^2 \mathbf{I}$, we employ three established techniques: L-curve \cite{hansen1999curve}, Generalized Cross-Validation (GCV) \cite{aster2018parameter, golub1979generalized, hastie2009elements, li1985stein}, and the $\chi^2$ method \cite{bennett1992inverse, mead2008parameter} to estimate the optimal scalar variance $\sigma_b^2$.

\subsubsection{L-curve}
The L-curve is a graphical technique used to select the optimal regularization parameter by plotting the norm of the regularized solution against the norm of the residual norm for a range of regularization parameters. The name `L-curve' originates from the characteristic L-shaped curve observed in linear problems. The optimal regularization parameter is the one that corresponds to the point of maximum curvature, or the `corner' of the L, where the trade-off between regularization and data misfit is balanced \cite{hansen1999curve}.

While the L-curve approach provides a visually intuitive approach for parameter selection, it is fundamentally heuristic and choice of the corner point lacks a statistical basis. Furthermore, the L-curve  may not always yield a well-defined corner, making its interpretation somewhat subjective \cite{johnston2000selecting, vogel1996non}. 

In the 3D-Var framework, if $\mathbf{B} = \sigma_b^2 \mathbf{I}$, the L-curve is obtained by plotting the regularization norm $\|\hat {\mathbf{x}} - \mathbf x _b\|_2$ against the data misfit $\|\mathbf{d} - \mathbf{H}\hat {\mathbf{x}}\|_2$. The optimal background error variance $\sigma _b ^{2}$ is selected at the point of maximum curvature.

\subsubsection{Generalized cross-validation (GCV)} The GCV method is a widely used approach for selecting the regularization parameter in Tikhonov regularization. It {assess} how well the regularized solution generalizes to new data by systematically leaving out individual observations and measuring the resulting prediction error \cite{aster2018parameter, golub1979generalized}. This approach ensures that the chosen regularization parameter effectively minimizes the trade-off between fitting the data and maintaining a smooth, stable solution, thereby enhancing generalization performance. The unified framework in \cite{tandeo2020review} identifies cross-validation as one of the methods that use the likelihood of the observations.

For 3D-Var, if $\mathbf{B} = \sigma_b^2 \mathbf{I}$ the GCV criterion is expressed as function of $\sigma _b$: 
\begin{equation}
    g(\sigma _b) =  \frac{M\|\mathbf{d} - \mathbf{H}\hat {\mathbf{x}}\|_2^2}{(\text{Tr}(\mathbf{I} - \mathbf{H} (\mathbf{H}^T \mathbf{R}^{-1}\mathbf{H} +  \mathbf{B}^{-1})^{-1} \mathbf{H}^T\mathbf{R}^{-1}))^2}~.
    \end{equation}
where $M$ is the number of observations. The optimal $\sigma _b ^{2}$ is found by minimizing the GCV function:
\begin{equation}
    {\sigma _b} = \arg\min_{\sigma _b} g(\sigma _b).
\end{equation}
\subsubsection{$\chi ^2$ method}
The $\chi^2$ method selects the regularization parameter based
on the assumption that the minimized cost function should follow a $\chi^2$ distribution with $M$ degrees of freedom, where $M$ is the number of observations. In the 3D-Var framework, the cost function is: \begin{equation} \mathcal{J}(\hat{\mathbf{x}}) = (\mathbf{d} - \mathbf{H}\mathbf{x}_b)^T(\mathbf{H}\mathbf{B}\mathbf{H}^T + \mathbf{R})^{-1}(\mathbf{d} - \mathbf{H}\mathbf{x}_b), \end{equation}
and the regularization parameter is chosen such that:
\begin{equation} \mathcal{J}(\hat{\mathbf{x}}) \approx M. \end{equation}

The $\chi^2$ method is a method of moments where the expected value of $\mathcal{J}(\hat{\mathbf{x}})$ is used to estimate the regularization parameter. As described in \cite{tandeo2020review}, to date, most methods for estimating covariances use differences between observations and projections of forecasts onto the observation space. Alternatively, the $\chi^2$ method uses the regularized difference e.g.,(\ref{A1}), since this functional follows a $\chi^2$ distribution. Bennett explains in \cite{bennett2005inverse} Chapter 2.3 that the regularized cost functional at the minimum follows a $\chi^2$ distribution while a more detailed proof is given in  Mead \cite{mead2008parameter}.  This holds in cases with non-Gaussian errors since under mild regularity conditions, the distribution of the cost function converges  to a $\chi^2_M$ distribution as $M \sim \infty$. An efficient algorithm for the $\chi ^2$ method was developed in \cite{Mead_2020}.

The regularization parameter selection techniques mentioned above can be extended to the non-isotropic case where $\mathbf{B}$ allows for spatial correlations and heterogeneity in the error structure. This setting involves multiple hyperparameters, such as variances and spatial correlation length scales, which define a more realistic and structured background error covariance. 

For example, Brezinski \cite{brezinski2003multi} extended the Generalized Cross-Validation (GCV) method to multi-parameter cases by defining a multi-parameter GCV function that accounts for interactions among several regularization terms. Mead \cite{mead2013discontinuous} used the $\chi^2$ principle in the context of estimating diagonal weighting matrices rather than scalar regularization parameters. These multi-parameter frameworks are particularly relevant when estimating dense or structured background error covariance matrices $\mathbf{B}$, where the scalar isotropic assumption is insufficient to capture the true error behavior.

\subsection{Weak constraint 4D-Var model error variance estimation}\label{subsec_2.2}
In weak-constraint 4D-Var, the evolution of a system state $\mathbf{x} $ accounts for uncertainties arising from both the model and observations. Assume the system is governed by a process model:
\begin{align}
\begin{aligned}
\mathbf{x}_i &= \mathbf{M}_i \mathbf{x}_{i-1} + \boldsymbol{\delta}_i, \quad i = 1, \dots, T  \\
\mathbf{d}_i &= \mathbf{H}_i \mathbf{x}_i + \boldsymbol{\epsilon}_i\\
\mathbf{x}_0 &= \mathbf{x}_b + \boldsymbol{\eta}
\end{aligned} \label{A5}
\end{align}
where $\mathbf{x}_i \in \mathbb{R}^N$ represents the state vector at time $t_i$ and $\mathbf{M}_i$ is the model operator that propagates the state from $t_{i-1}$ to $t_i$. The term $\boldsymbol{\delta}_i \sim \mathcal{N}(\mathbf{0}, \mathbf{Q}_i)$ represents model errors. Observations $\mathbf{d}_i \in \mathbb{R}^M$ are obtained through the observation operator $\mathbf{H}_i$, with measurement errors $\boldsymbol{\epsilon}_i \sim \mathcal{N}(\mathbf{0}, \mathbf{R}_i)$. The initial state $\mathbf{x}_0$ is estimated from the background state $\mathbf{x}_b$, where the background error $\boldsymbol{\eta} \sim \mathcal{N}(\mathbf{0}, \mathbf{B})$.

We consider the first time step which is similar to 3D-Var, the state $\mathbf{x}_1$ is estimated by minimizing the following cost function:
\begin{align}
    \begin{aligned}
         \mathcal J(\mathbf{x}_0, \mathbf x_1) &= \frac{1}{2} \begin{bmatrix}
             (\mathbf{x}_0 - \mathbf{x}_b) & (\mathbf{x}_1 - \mathbf{M}_1 \mathbf{x}_{0})
         \end{bmatrix} \mathbf{C}^{-1}\begin{bmatrix}
             (\mathbf{x}_0 - \mathbf{x}_b) & (\mathbf{x}_1 - \mathbf{M}_1 \mathbf{x}_{0})
         \end{bmatrix}^T
         \\ & \quad + \frac{1}{2} (\mathbf{d}_1 - \mathbf{H}_1\mathbf{x}_1)^T \mathbf{R}_1^{-1} (\mathbf{d}_1 - \mathbf{H}_1\mathbf{x}_1), 
    \end{aligned} \label{A6}
\end{align}
and the covariance matrix $\mathbf{C}$ is given by
\[
\mathbf C = \begin{bmatrix} 
\mathbf B & \mathbf B \mathbf M_1^T \\ \\
\mathbf M_1 \mathbf B & ~~~\mathbf M_1 \mathbf B \mathbf M_1^T + \mathbf Q_1 
\end{bmatrix}.
\]
Assuming the background error covariance $\mathbf{B}$ is known, the model error covariance $\mathbf{Q}_1$ can be estimated using regularization parameter selection methods, similar to the approach in 3D-Var (see \cref{subsec_2.1}). Note that in the first time step, this means the unknown covariance is searched for in a $2N \times 2N$ dimensional space. When we proceed through time in this manner, the unknown covariance is searched for in a space of dimension $(T+1)N \times (T+1)N$. For reference in our experiments in \cref{sec:4} with $49$ data points, the search space at a single time step is of dimension $49 \times 49$ while the global 4D-Var approach over all time steps involves searching through a significantly larger space of $89400 \times 89400$. Although the global  4D-Var approach captures all information in time, it poses substantial computational challenges. To  make this approach feasible, we use the representer method (see \cref{sec:3}) to obtain the optimal estimate efficiently. The representer method reduces the search space to the data space (e.g $49 \times 49$ in our experiments).

%%%%%%%%%%%%%%%%%%%%%%%%%%%%%%%%%%%%%%%%%%%%%%%%%%%%%%%%%%%%%%%%%%%%%%%

\section{Weak Constraint 4D-Var with Representers} \label{sec:3}
In \cref{sec:2}, we used a discrete matrix-based representation $\mathbf{M}_i$ for the process model (\ref{A5}). Here we consider a continuous  partial differential equation (PDE) which allows the proposed framework to be applied to a broader range of numerical models that may not be clearly expressed as matrix multiplications. We choose the governing equation: 
\begin{equation}
    \begin{aligned}
        \frac{\partial {q}}{\partial t} + L[q(\mathbf{x},t)] & = Q(\mathbf{x},t)+ f(\mathbf{x},t) & \text{for } \mathbf{x} \in \Omega, \quad t\in[0,T] \\
        q(\mathbf{x},0) & = I(\mathbf{x}) + i(\mathbf{x}) \\
        q(\mathbf{0},t) & = B(t) + b(t)
    \end{aligned} \label{3.1}
\end{equation}
where $q(\mathbf{x},t)$ represents the state of the system, the operator  ${L}[q(\mathbf{x},t)]$ denotes the model dynamics which may be linear or nonlinear, $Q(\mathbf{x},t)$ represents the external forcing, $I(\mathbf{x})$ represents the initial state of the system and $B(t)$ represents the boundary conditions. We assume that there are errors $f(\mathbf x,t)$, $i(\mathbf x)$ and $b(t)$ in the model dynamics (or external forcing), initial and boundary conditions with covariances  $C_f(\mathbf{x},t)$, $C_i(\mathbf{x})$ and $C_b(t)$ respectively. These errors are assumed to be unbiased and mutually uncorrelated.  

We assume that imperfect data are available at a limited number of points $(\mathbf{x}_m,t_m)$, collected at $M$ points in space and time. These observations are related to the “true” state $q(\mathbf{x},t)$ by
\begin{equation}
   {d}_m = H_m{q} (\mathbf{x}_m,t_m)+{\epsilon}_m~~~~~~~~~~~m=1,2,\cdots, M  \label{3.2}
\end{equation}
where ${\epsilon} _m \sim \mathcal N (0,\sigma_m ^2)$ is the  measurement error with error variance $\sigma_m ^2$ and $H_m$ is an observation operator that transforms $q(\mathbf{x},t)$ into observation equivalents.

To obtain the optimal state estimate $\hat{q}(\mathbf x , t)$ that best fits the model and the observations, we minimize a weighted least-squares cost function $\mathcal{J}$:
 \begin{equation}
     \hat{q}(\mathbf x , t) = \arg \min _{q}  \mathcal{J}[q] \label{3.3}
 \end{equation}
 where
\begin{align}
    \begin{aligned}
        \mathcal{J}[q]={} & \int_{0}^{T}\int_{\Omega} \int_{0}^{T}\int_{\Omega}  f(\mathbf x , t) {C_f ^{-1}(\mathbf{x},t,\mathbf{x}',t') }f(\mathbf x' , t') d\mathbf{x}' dt' d\mathbf{x}dt   \\& + \int_{\Omega} \int_{\Omega} i (\mathbf{x}) {C_i ^{-1} (\mathbf{x},\mathbf{x}')} i(\mathbf{x}') d\mathbf{x}' d\mathbf{x}  \\& \quad + \int_{0}^{T} \int_{0}^{T} b(t) {C_b ^{-1}(t,t')} b(t') dt' dt +   \sum_{m=1}^{M} \sigma_m^{-2} \epsilon _m^2.
    \end{aligned} \label{3.4}  
\end{align}

We seek to find the optimal state estimate $\hat{q}(\mathbf x,t)$ that corresponds to the smallest values of the errors $f(\mathbf {x},t)$, $i(\mathbf {x})$, $b(t)$ and \( \epsilon _m\). Due to the lack of observations at all points in space and time, the error fields $f(\mathbf {x},t)$, $i(\mathbf {x})$ and $b(t)$  are undetermined. However with appropriate weights for these errors, there is a unique optimum $\hat{q}(\mathbf {x},t)$ that minimizes \( f(\mathbf {x},t) \), \( i(\mathbf {x}) \), \( b(t) \), and \(  \epsilon _m \) in a weighted least-squares sense. Given that \( \mathcal{J} \) is non-negative and quadratic in \( q(\mathbf {x},t) \), we find the global minimum using calculus of variations. This results in coupled Euler-Lagrange equations (\ref{3.5}). 

The result in this work applies to any operator $L$. However, the Euler-Lagrange (E-L) equations will change depending on the type of boundary conditions used. Therefore, we omit boundary condition errors for generalization purposes. Assuming that the operator $L$ has been linearized, the Euler-Lagrange equations are
\begin{align}
    \begin{aligned}
        -\frac{\partial \lambda}{\partial t} -  L^T\lambda (\mathbf{x},t) &= - \sum_{m=1}^{M} \sigma_m^{-2} (H_m\hat{q}(\mathbf{x}_m,t_m)- d_m)\delta (\mathbf{x} - \mathbf{x}_m)\delta (t-t_m) \\
        \lambda (\mathbf{x},T) & = 0\\
        \frac{\partial \hat{q}}{\partial t} + L\hat{q} (\mathbf{x},t)   
        & = Q(\mathbf x,t) + {C_f \bullet \lambda (\mathbf x,t)}\\
        \hat{q}(\mathbf{x},0) & =  I(\mathbf{x}) + {C_i\circ \lambda(\mathbf x,0)} 
    \end{aligned}\label{3.5}
\end{align}
where $\delta(\mathbf{x})$ is the Dirac delta function, the adjoint variable $\lambda (\mathbf x,t)$ is defined as the weighted residual: 
\begin{equation}
    \lambda (\mathbf x,t) = \int_{0}^{T}\int_{\Omega} C_f ^{-1}(\mathbf{x},t,\mathbf{x}',t') f(\mathbf x' , t') d\mathbf{x}' dt',\label{3.6}
\end{equation}
while the covariance multiplication with the adjoint is the convolution expressed as  
\begin{equation}
   Cf \bullet \lambda (\mathbf x,t)  = \int_{0}^{T} \int_{\Omega}  C_f(\mathbf{x},t,\mathbf{x}',t') \lambda (\mathbf x',t') d\mathbf{x}' dt', \label{3.7}
\end{equation}
and 
\begin{equation}
   C_i\circ \lambda(\mathbf x,0)  = \int_{\Omega}  C_i(\mathbf{x},\mathbf{x}') \lambda (\mathbf x',0) d\mathbf{x}'.\label{3.8}
\end{equation}

The coupled system of equations (\ref{3.5}) requires $\lambda$ to be solved backward in time and the optimal estimate $\hat{q}$ forward in time. We decouple this linear system of equations by using the representer method. 

\subsection{Representer method} \label{subsec:3.1}
The representer method for 4D-Var \cite{bennett2005inverse} is based on the representer theorem \cite{wahba1990spline, wahba2019representer, scholkopf2001generalized}. Here, we present an alternative derivation of the method than in \cite{bennett1992inverse}. Our derivation shows how the method is explicitly grounded in approximation theory and its formulation lies within the reproducing kernel Hilbert space (RKHS) framework that has gained recent popularity in learning problems.  Consider (\ref{3.4}) rewritten as a regularized functional:
\begin{equation}
\mathcal J[q] = \boldsymbol{\epsilon}^T \mathbf{C}^{-1}_{\epsilon}\boldsymbol{\epsilon} + \mathcal{R}(q),\label{3.9}
\end{equation}
 where $\boldsymbol{\epsilon}$ represents the data misfit with error covariance  $\mathbf{C}_{\epsilon} = \diag(\sigma _1^2,\sigma _2^2, \dots, \sigma _M^2)$ and $\mathcal{R}(q)$ is the regularization term derived from the governing PDE. The goal is to find the optimal state $\hat{q}$ by minimizing $\mathcal{J}[q]$ over an infinite-dimensional space $\mathcal{H}$. The function space $\mathcal{H}$ can be decomposed as \( \mathcal{H} = \mathcal{H}_0 \oplus \mathcal{H}_1 \), where \( \mathcal{H}_0 = \{ q : \mathcal{R}(q) = 0 \} \) is a finite-dimensional subspace spanned by basis functions \( \phi_1, \dots, \phi_N \) and \( \mathcal{H}_1 \) is a RKHS with reproducing kernel \( \Gamma \). Define representer functions \( r_m(\mathbf{x}, t) = \Gamma(\mathbf{x}, t, \mathbf{x}_m, t_m) \). The representer theorem states that the optimal estimate can be expressed as:
\begin{equation}
    \hat{q}(\mathbf{x}, t) = \sum_{n=1}^{N} c_n \phi_n(\mathbf{x}, t) + \sum_{m=1}^{M} \beta_m r_m(\mathbf{x}, t). \label{3.10}
\end{equation}
Let $q_F(\mathbf{x}, t)$ denote the solution to the error-free model in (\ref{3.1}). Since by construction $\mathcal{R}(q_F) = 0$ in cost functionals (\ref{3.4}) and (\ref{3.9}), it follows that $q_F \in \mathcal{H}_0$. Therefore, the optimal estimate simplifies to:
\begin{equation}
    \hat{q}(\mathbf{x}, t) = q_F(\mathbf{x}, t) + \sum_{m=1}^{M} \beta_m r_m(\mathbf{x}, t)\label{3.11}.
\end{equation}
Not only has the representer theorem reduced searching for the optimal estimate in infinite-dimensional space of the PDE to finite-dimensional space but the size of the finite-dimensional search space is $M$, the number of data. This is particularly advantageous in data assimilation where number of data is significantly smaller than the number of state estimates $N$.

Next, we describe the approach for finding scalar coefficients $\beta _m$ and representer functions $r_m (\mathbf{x},t)$. Define the adjoint variable $\lambda (\mathbf{x},t)$ as a linear combination of adjoint representer functions $\alpha_m(\mathbf{x}, t)$:
\begin{equation}
\lambda (\mathbf{x},t) = \sum_{m=1}^{M} \beta_m \alpha_m(\mathbf{x}, t).
\end{equation}

Substituting (\ref{3.11}) into the  forward E-L equations we get: 
\begin{align}
(F_m)\begin{cases}
        \frac{\partial r_{m}}{\partial t} +  L r_{m} (\mathbf{x},t) = C_f \bullet  \alpha_m(\mathbf{x},t)\\
        r_m(\mathbf{x},0) = C_i\circ \alpha_m(\mathbf{x},0)
    \end{cases},\label{8}
\end{align}
and we define the adjoint representers to be the Green's function for the backward E-L equations, i.e.
 \begin{align}
    (B_m)\begin{cases}
        -\frac{\partial \alpha_m}{\partial t} - L^T \alpha_m (\mathbf{x},t) = H_m^T\delta(\mathbf{x}-\mathbf{x}_{m})\delta(t-t_{m})\\
        \alpha_m(\mathbf{x},T)=0
    \end{cases}.\label{9}
\end{align}
The representer functions $r_m (\mathbf{x},t)$ are obtained by first solving the for $\alpha_m (\mathbf{x},t)$ backward in time, then using it to solve for $r_m (\mathbf{x},t)$ forward in time for $1\leq m \leq M$. The scalar coefficients are found by substituting (\ref{3.11}) into (\ref{3.5}). This gives $M$ equations for the unknown representer coefficients $\boldsymbol{\beta} = [\beta _1, \beta _2, \cdots, \beta _M] ^T$ and 
\begin{equation}
     \boldsymbol{{\beta}} = (\mathbf{R}+ \mathbf{C}_{\epsilon})^{-1} (\mathbf{d} - {\mathbf{q}_F}_m), \label{10}
    \end{equation}
where ${\mathbf{q}_F}_m = [{q}_F(\mathbf x _1,t_1),{q}_F(\mathbf x _2,t_2),
\cdots, {q}_F(\mathbf x _M,t_M)
]^T$, $\mathbf d = [d_1, d_2, \cdots d_M]^T$ and 
$\mathbf{R}\in \mathrm{R}^{M \times M}$ is the representer matrix evaluated data locations, i.e.
$\mathbf{R}_{m_1m_2} =  r_1(\mathbf{x}_{m_2},t_{m_2})$. Substituting for $\beta _m$ in (\ref{3.11}), yields the final expression for the optimal estimate:
\begin{equation}
    \hat{q}(\mathbf{x},t) = q_F(\mathbf{x},t)+\mathbf{h}^T \mathbf{P}^{-1}\mathbf r(\mathbf{x},t), \label{3.15}
\end{equation}  
where $\mathbf{P} = \mathbf{R}+ \mathbf{C}_{\epsilon}$, $\mathbf h = \mathbf{d} - {\mathbf{q}_F}_m$ and $\mathbf r(\mathbf{x},t) = [
    r_1(\mathbf x, t), r_2(\mathbf x, t), \cdots, r_M(\mathbf x, t)]^T$.

Note that the analytical expression (\ref{3.15}) for the optimal estimate for 4D-Var with representers resembles (\ref{2.2}) for the 3D-Var formulation. In particular, the dimension of the space we need to search for covariances is $M \times M$ for both 3D-Var and 4D-Var with representers. The representer method is a significant decrease in cost especially when $N >> M$.
As we mentioned in \cref{subsec_2.2}, a 4D-Var approach without representers would result in searching for covariances in space of dimension $(T+1)N \times (T+1)N$.

\subsection{Model error covariance estimation}\label{sec:32}
The variational formulation is expressed as: 

\begin{align}
    \begin{aligned}
        \hat{q} ={}& \arg \min_{q} \bigg\{ \boldsymbol{\epsilon}^T \mathbf{C}^{-1}_{\epsilon}\boldsymbol{\epsilon} + \int_{0}^{T}\int_{\Omega} \int_{0}^{T}\int_{\Omega}  f(\mathbf x , t) C_f ^{-1}(\mathbf{x},t,\mathbf{x}',t') f(\mathbf x' , t') d\mathbf{x}' dt' d\mathbf{x}dt \\&    + \int_{\Omega} \int_{\Omega} i (\mathbf{x}) C_i ^{-1} (\mathbf{x},\mathbf{x}') i(\mathbf{x}') d\mathbf{x}' d\mathbf{x}  + \int_{0}^{T} \int_{0}^{T} b(t) C_b ^{-1}(t,t') b(t') dt' dt.
    \end{aligned}\label{3.17}
\end{align} 
We assume that the data error covariance $\mathbf{C}_{\epsilon}$
is known since this can often be derived from the characteristics of the data, such as instrument errors or inferred using standard error models. Similarly, the initial condition and boundary condition error covariances, $C_i(\mathbf{x})$ and $C_b(t)$ are assumed to be known or treated as exact. Under this setup, the formulation in (\ref{3.17}) can be viewed as a Tikhonov regularization problem, where the regularization term reflects weakly constrained model dynamics and the inverse model error covariance $C_f^{-1}(\mathbf{x},t)$ as the regularization matrix.  

\subsubsection{Isotropic model error covariance}\label{subsec_3.2.1}
We begin with the isotropic case, where the model error covariance is assumed to be of the form: 
\begin{equation} C_f = \sigma_f^2 \mathbf{I}, \label{3.18} \end{equation} where $\sigma _f^2$ is the model error variance and $\mathbf I$ is an identity matrix. In this case, the covariance estimation problem reduces to estimating a single parameter, $\sigma _f^2$.

\begin{figure}[H]
    \centering
    \begin{subfigure}[b]{0.35\textwidth}
        \includegraphics[width=\textwidth]{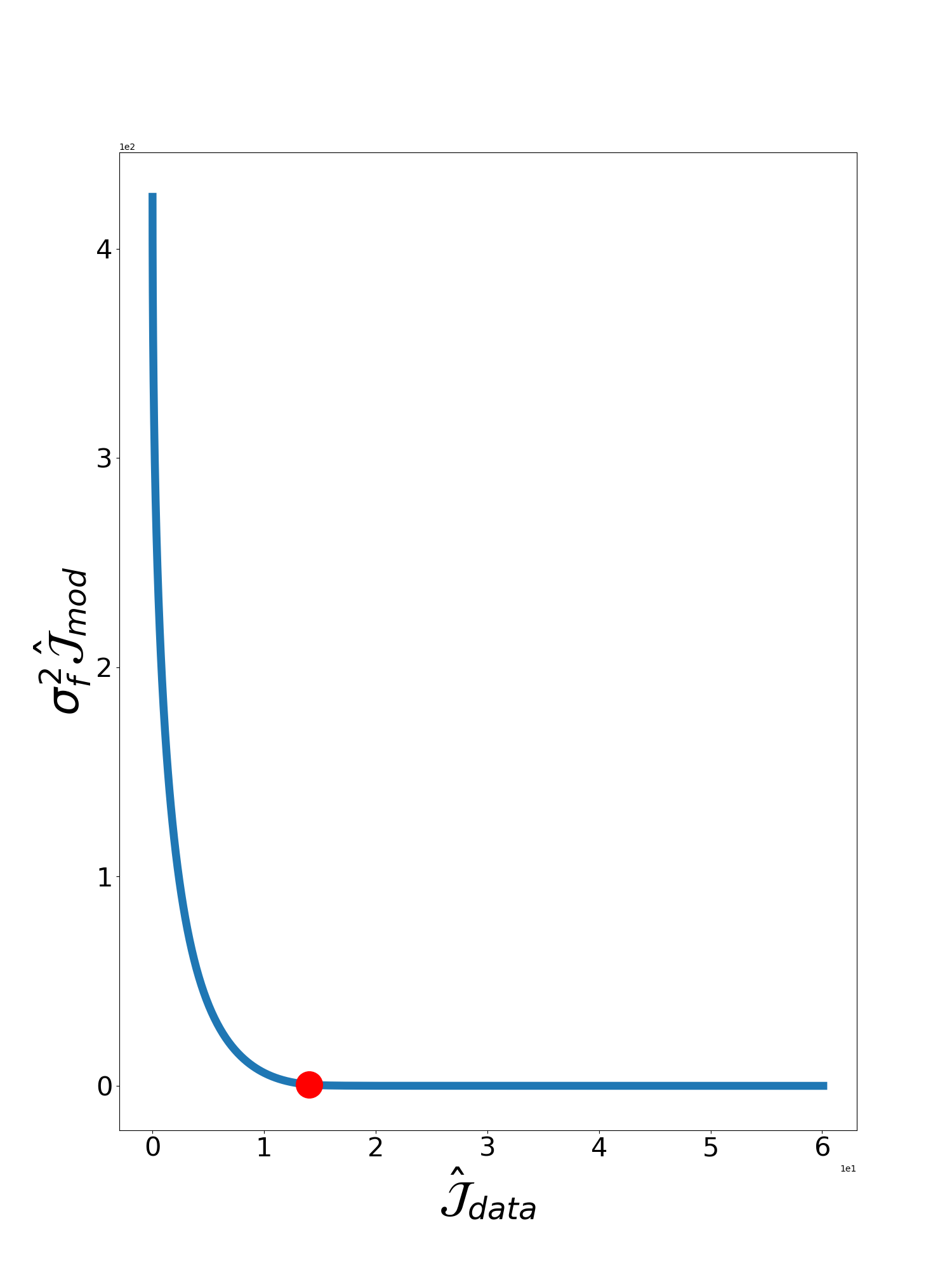}
        \caption{L-curve.}
    \end{subfigure}
    \begin{subfigure}[b]{0.35\textwidth}        \includegraphics[width=\textwidth]{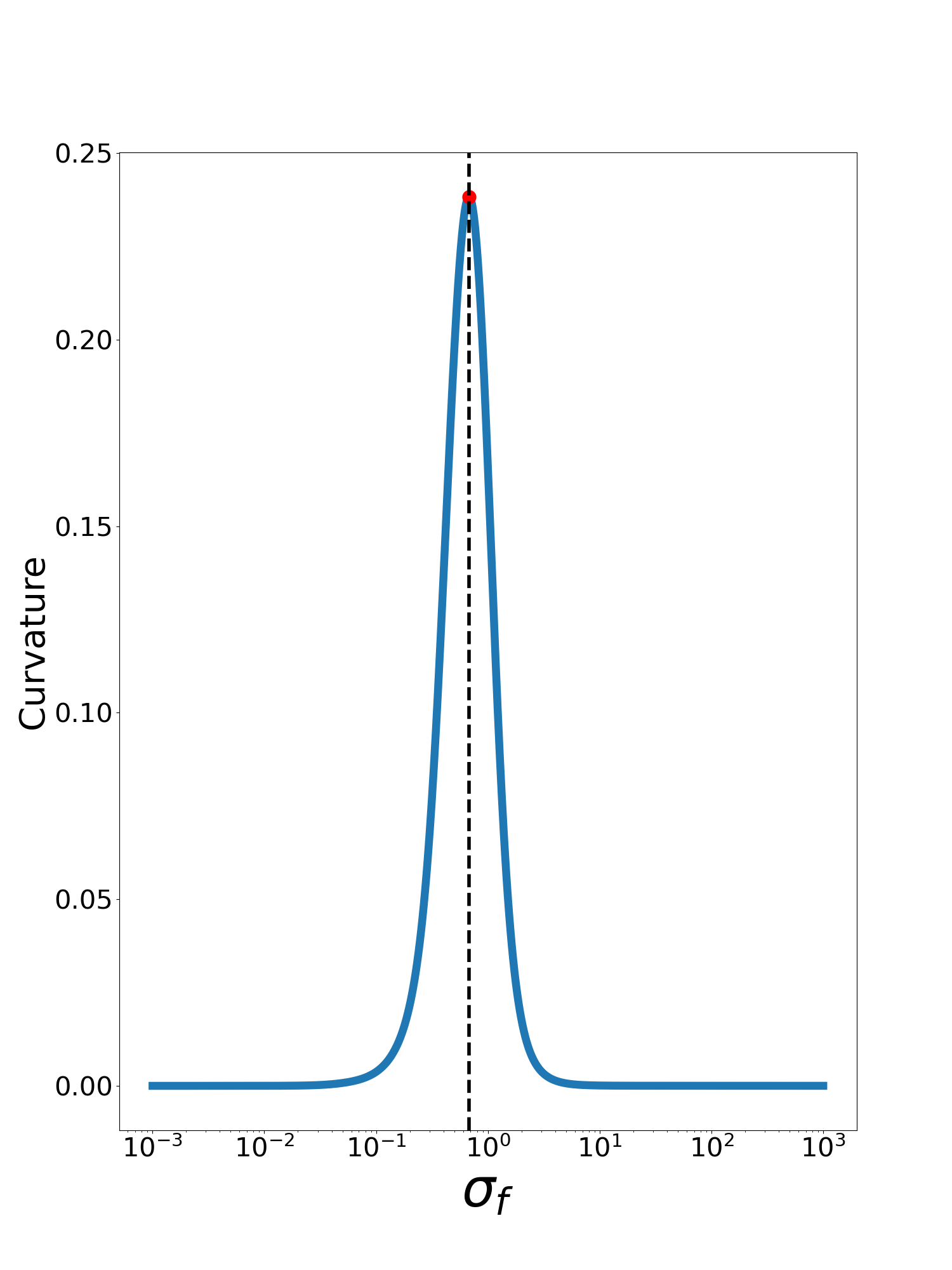}
        \caption{Curvature of the L-curve}
    \end{subfigure} 
    \caption{Illustrative example of model error variance estimation for 4D-Var with representers using the L-curve and it's curvature. The red dot indicates the point of maximum curvature, corresponding to the optimal $\sigma_f$.}\label{fig:1}
\end{figure}

\textbf{L-curve:} The L-curve is obtained by plotting the regularization norm $\sigma _f^{2} \mathcal{\hat{J}}_{\text{mod}}$ against the data misfit $\mathcal{\hat{J}}_{\text{data}}$ as shown in \cref{fig:1} (a). The expressions for the data misfit and regularization norm are derived using \cref{lemma1} and \cref{lemma2}, respectively whose proofs are shown in \cref{appendix_A}.

\cref{fig:1}(a) shows a typical L-curve, where the regularization norm (model error) is plotted against the data misfit for different values of $\sigma_f^2$. The red dot marks the point of maximum curvature, often indicating the optimal regularization parameter. \cref{fig:1}(b) provides additional insight by plotting the curvature of the L-curve as a function of $\sigma_f$, with the peak corresponding to the optimal $\sigma_f^2$. However, as noted by Hansen \cite{hansen1999curve}, the L-curve corner doesn't always align with the maximum curvature, especially when the regularization norm doesn't rise sharply for regularization parameters beyond the optimal $\sigma_f^2$. 

\textbf{Generalized cross validation method:} Extending the GCV concept in \cref{subsec_2.1} to weak constraint 4D-Var with representers, we derive a GCV function that can be used to estimate the model error variance  $\sigma _f^{2}$. 
\begin{theorem} Let $\mathbf{C}_{\epsilon} = \diag(\sigma _1^2,\sigma _2^2, \dots, \sigma _M^2)$, $w_m = \sigma _m^{-2} $ for $m=1,2,\cdots,M$ and assume that $C_i$ and $C_b$ are specified. The generalized cross validation function for weak constraint 4D-Var with representers is \label{thm:4}
 \begin{equation}
    g(\sigma _f) = \frac{1}{M} \sum_{k = 1}^M w_k\left(\frac{\hat{q} (\mathbf x_k,t_k) - {d}_k}{1-(\mathbf{R} \mathbf{P}^{-1})_{kk}}\right)^2.\label{gcv_22} 
\end{equation}
\end{theorem} 
The proof of theorem is shown in \cref{appendix_C}. The optimal model error variance $\sigma_f^2$ is the one which minimize (\ref{gcv_22}). This method evaluates how well the model and observational data are balanced, minimizing the prediction error by systematically adjusting $\sigma_f^2$. 
 
\textbf{$\boldsymbol{\chi ^2}$ method:} In this approach, the optimal value of $\sigma _f^{2}$ is estimated such that the minimized cost function $\hat{\mathcal{J}}$ in \cref{theorem:3} follows a  $\chi ^2$ distribution with $M$ degrees of freedom. i.e. 
\begin{equation}
   \hat{\mathcal{J}} \equiv \mathcal{J}[\hat{q}] = \mathbf{h}^T \mathbf{P}^{-1} \mathbf{h} \sim \chi ^2 _M.
\end{equation}
The optimal value of $\sigma_f^2$ is obtained by solving:
\begin{equation}
   \hat{\mathcal{J}}(\sigma_f) = \mathbf{h}^T \mathbf{P}^{-1} \mathbf{h} =  M.
\end{equation}
It  has been demonstrated in in \cite{Mead_2020}
that for a convergent optimization algorithm, the minimized cost function is monotonically decreasing as a function of $\sigma _f^2$.

\subsubsection{Non-Isotropic model error covariance}\label{subsec_3.2.2}
To incorporate spatial and temporal correlation in model errors, we consider a non-isotropic covariance of the form:
 \begin{equation} 
 C_f(\mathbf{x}, t, \mathbf{x}', t') = \sigma_f^2 \exp\left(-\frac{|\mathbf{x} - \mathbf{x}'|^2}{2l_f^2}\right) \exp\left(-\frac{|t - t'|}{\tau_f}\right), \label{3.19} 
 \end{equation} 
where $\sigma_f^2$ is the model error variance, $l_f$ is the spatial correlation length scale and $\tau_f$ is the temporal correlation scale. This covariance structure is separable in space and time and it is Gaussian in space, and corresponds to a first-order Markov process in time. This covariance form has been widely used in previous representer-based data assimilation studies \cite{bennett1996generalized, ngodock2005efficient}.

Unlike prior work where the hyperparameters $\sigma_f^2$, $l_f$ and $\tau_f$, are fixed apriori, we develop a framework to estimate them using multiple regularization parameter selection methods i.e the GCV \cite{brezinski2003multi} and the $\chi^2$ method \cite{mead2013discontinuous}. In this case we do not use the L-hypersurface, as it can be computationally intensive given the increased dimensionality of the hyperparameter space.

\textbf{Multi-parameter GCV}
We generalize the GCV function to handle multiple hyperparameters, i.e., 
 \begin{equation}
    g(\sigma_f^2, l_f \tau_f) = \frac{1}{M} \sum_{k = 1}^M w_k\left(\frac{\hat{q} (\mathbf x_k,t_k) - {d}_k}{1-(\mathbf{R} \mathbf{P}^{-1})_{kk}}\right)^2.
\end{equation}\label{gcv_3}
The optimal values of $\sigma_f^2$, $l_f$ and $\tau_f$ are obtained by minimizing this function.

\textbf{Multi-parameter $\boldsymbol{\chi ^2}$ method:}
Similarly, the $\chi ^2$ method approach is extended by solving:
\begin{equation} \hat{\mathcal{J}}(\sigma_f^2, l_f \tau_f) = \mathbf{h}^T \mathbf{P}^{-1} \mathbf{h} = M, \end{equation} 
with respect to all three hyperparameters simultaneously.

\section{Numerical Experiments}\label{sec:4}
In this section, we present numerical experiments designed to evaluate the performance of our proposed regularization parameter selection framework for weak constraint 4D-Var data assimilation. We estimate the state of wildfire smoke PM$_{2.5}$ concentration, $q(\mathbf{x},t)$, using a 1D transport equation and simulated observational data. The primary goal is to find model error covariance estimates that properly balance the influence of the model and observational errors. We consider both isotropic and non-isotropic formulations of the model error covariance and apply the regularization parameter selection techniques described in \cref{sec:32} to estimate the corresponding hyperparameters: $\sigma_f^2$ for the isotropic case and $\sigma_f^2, l_f, \tau_f$ for the non-isotropic case.

The model setup which includes the source term, initial and boundary conditions, first guess, and experiment parameters is identical across the two cases to ensure fair comparisons. The key differences between the two cases lie in the data used and the spatial-temporal resolution of the discretized domain. For the isotropic case, we use a spatial grid of 200 points and 445 time steps, resulting in a state dimension of $N=89,000$. We generate a dataset $\mathbf D \in \mathbb{R} ^{49 \times 500}$ to allow for statistical analysis across 500 independent assimilations. For the non-isotropic case, we reduce the grid resolution to 50 spatial points and 112 time steps (state size $N=5763$) to reduce the computational cost. In this case, we use only one column of data i.e., $\mathbf d \in \mathbb{R} ^{30 \times 1} $. Although our current implementation is limited to smaller grids, future work will extend this approach to larger-scale, high-resolution cases. In both cases, we assume that the model error arises entirely from uncertainties in the source term, while the initial and boundary conditions are taken to be exact.
\subsection{Model and data setup}
\subsubsection{Transport model}
The PM$_{2.5}$ concentration $q(x,t)$ is modeled by a 1D transport equation with an exponential source term $Q(x,t)$ representing wildfire emissions:
\begin{align}
    \begin{aligned}
        \frac{\partial {q}}{\partial t} + {u} \frac{\partial {q}}{\partial x} & = Q({x},t) & \text{for } {x} \in [30,45], \quad t\in[0,20] \\
        q({x},0) &= 0 \\
        q({0},t) &= B(t)  \end{aligned}\label{C4_1}
\end{align}
where $u=1$ is a constant wind field. Since the main focus of this paper is estimating the error variance of the model dynamics, we assume that the initial and boundary conditions are exact. We solve the transport model using the upwind finite volume method \cite{leveque2002finite} described in \cref{appendix_B}.

The source term $Q(x,t)$ is designed to model the emission of PM$_{2.5}$ from wildfire sources and is defined as the sum of two Gaussian functions. Each function represents a distinct wildfire source, characterized by its emission strength, spatial and temporal decay rates, and the location. Specifically, the source term is given by:
\begin{equation}
 Q(x,t) = S_0 \exp(- \alpha_0 (x-x_0)^2-k_0t) + S_1 \exp(- \alpha_1 (x-x_1)^2-k_1t), \label{C4_3}
\end{equation}
where $S_0$ and $S_1$ denote the initial emission strengths of the two sources, while $\alpha_0$, $\alpha_1$ are the spatial decay rates, controlling the rate at which the PM$_{2.5}$ concentration diminishes with distance from the source locations $x_0 = 33$ and $x_1 = 40$. The parameters $k_0$ and $k_1$ represent the temporal decay rates, dictating how the emission intensity decreases over time. This formulation allows the model to capture the temporal and spatial dynamics of PM$_{2.5}$ dispersion resulting from multiple wildfire sources.

In this experimental setup, the model error is completely defined by parameters in the source term, we note that this problem could be set up as strong constraint 4D-Var if the state estimation is augmented with estimation of the source term parameters.

\subsubsection{Observations}
In all experiments, we generate synthetic observations by randomly sampling $M=49$ spatial-temporal points for the isotropic case and $M=30$ for the non-isotropic case from a uniform distribution over $[30,45] \times [0,20]$. The simulated observations are generated using
\begin{equation} {d}_m = H_m q(x_m, t_m) + \boldsymbol{\epsilon}_m, \quad \boldsymbol{\epsilon}_m \sim \mathcal{N} (0,\sigma_m^2), \label{data2} \end{equation}
where $q(x_m,t_m)$ represents the true concentration at each sampled location, and $\boldsymbol{\epsilon}_m$ denotes measurement noise. Since the observation point  $(x_m,t_m)$ may not lie directly on the computational grid, the operator $H_m$ is an interpolation operator that maps the continuous state $q(x,t)$ onto the observation location. The noise variance $\sigma _m^2$ is proportional to the true concentration and is given by:
\begin{equation} 
\sigma_m = \sigma q(x_m,t_m). 
\end{equation}
where $\sigma$ is a specified noise level that varies across experiments.  This ensures that regions with higher PM$_{2.5}$ concentrations exhibit greater variability and captures the inherent uncertainty and measurement errors typical of environmental monitoring systems.

To account for sampling variability and ensure statistical robustness, we generate a dataset $\mathbf{D} \in \mathbb{R}^{49 \times 500}$ for each experiment. Specifically, we run equation (\ref{data2}) $10^5$ times to compute the mean and standard deviation of the root mean-squared error (RMSE) between the simulated observations and the true concentrations $q(x_m, t_m)$, obtained by solving the PDE (\ref{C4_1}). Based on these statistics, each column $\mathbf{d} \in \mathbb{R}^{49}$ is selected such that its RMSE lies within the range $\text{mean}(RMSE) \pm \text{std}(RMSE)$.

\subsubsection{First guess}
The first guess $q_F (x,t)$ for the PM$_{2.5}$ concentration is obtained by solving the same transport equation used to generate the true solution, but with perturbed parameter values in the source term. These perturbations are applied to the spatial decay rates, and temporal decay rates of the wildfire sources, introducing discrepancies between the first guess and the true solution. Specifically, the source term for the first guess is defined with slightly different parameters ${\alpha_F}_0$, ${\alpha_F}_1$, ${k_F}_0$ and ${k_F}_1$ compared to the true solution with ${\alpha_F}_i \sim \mathcal{N} (\alpha_i,\sigma _{\alpha_i}^2)$ and ${k_F}_i \sim \mathcal{N} (k_i,\sigma _{k_i}^2)$  for $i =0,1$. i.e.
\begin{equation}
 Q_F(x,t) = S_0 \exp(- {\alpha_F}_0 (x-x_0)^2-{k_F}_0t) + S_1 \exp(- {\alpha_F}_1 (x-x_1)^2-{k_F}_1t) \label{C4_5}.
\end{equation}
This setup represents a perfect-model twin experiment in which the only source of model error arises from the stochastic perturbations in the source term parameters, not from structural differences in the model.
The first guess serves as a baseline from which the data assimilation process begins, highlighting the need for adjustments based on observational data to achieve more accurate state estimates.

\subsection{Experiment setup} 
The experimental setup is designed to evaluate the performance of model error estimation in weak constraint 4D-Var data assimilation under different data noise levels and errors in the dynamical model. The experiments are classified into two categories based on whether the simulated observational data or first guess is trusted more. In the first category, experiments 1 and 2 in \cref{tab:1}, the first guesses are more accurate, representing situations where the model is more trusted than the data. In the second category, experiments 3 and 4 in \cref{tab:1}, the simulated observational data is closer to the true PM$_{2.5}$ solution reflecting scenarios where the data are more reliable than the model.
\FloatBarrier
\begin{table}[tbhp]
\centering
\footnotesize
\renewcommand{\arraystretch}{1.2}
% \resizebox{\textwidth}{!}{%
\begin{tabular}{|c|c|c|c|c|c|c|c|c|c|} 
\hline
\textbf{Expt} & \textbf{BC} &  $S_1$ &  $k_1$  & $\alpha _1$ & $\sigma$ & $\sigma _{k_0}$ & $\sigma _{k_1}$ & $\sigma _{\alpha _0}$ & $\sigma _{\alpha _1}$ \\
\specialrule{1.2pt}{0pt}{0pt}
1 & periodic & 0 & 0  & 0 & 0.7  & 0.2 & 0 & 0.2  & 0 \\ \hline
2 & no flux & 50  &  0.25  & 5 & 0.6 & 0.2 & 0.2 & 0.2 &  0.2 \\ \hline
3 & periodic  & 0  & 0  & 0 & 0.3  & 0.5 & 0  & 0.7 & 0  \\ \hline
4 & no flux & 50 & 0.25  & 5 & 0.2 &  0.6 & 0.5  & 0.5 & 0.5 \\ 
\specialrule{1.2pt}{0pt}{0pt}
\end{tabular}%
% }
\caption{Parameters and their corresponding standard deviations used in numerical experiments (Expt) 1–4.}
\label{tab:1}
\end{table}

\Cref{tab:1} shows the parameter values used in the experiments, including boundary conditions(BC), source emission strength, decay constants, noise levels, and perturbation magnitudes for the first guess. The parameters for the first source of PM$_{2.5}$ emissions in (\ref{C4_3}) are held constant across all experiments with $S_0=100$, $k_0 = 0.5$, and $\alpha_0 = 10$. Perturbations are applied to the parameters $k_0$, $k_1$, $\alpha_0$, and $\alpha_1$ in the first guess to introduce discrepancies between the model and the true solution, with $\sigma_{k_0}$ and $\sigma_{\alpha_0}$ corresponding to the perturbations in the temporal and spatial decay rates of the first source, and $\sigma_{k_1}$ and $\sigma_{\alpha_1}$ corresponding to the second source. $\sigma$ represents the noise level applied to the observational data.

\subsection{ Isotropic case: $C_f = \sigma _f^2 \mathbf I$}

\subsubsection{Estimation of model error variance $ \sigma _f^2$}
We use the L-curve, generalized cross-validation (GCV), and \texorpdfstring{$\chi^2$}{chi-squared} methods introduced in \cref{subsec_3.2.1} to estimate the model error variance \texorpdfstring{$\sigma_f^2$}{sigma-f-squared}. Each method is applied to all 500 columns of \texorpdfstring{$\mathbf{D}$}{D} to obtain an ensemble of estimates.

For the L-curve method, the values of ${\sigma _f} ^2$ at the corner of the L-curve were not necessarily the points of maximum curvature. As discussed in \cite{hansen1999curve}, the L-curve criterion only pinpoints the optimal regularization parameter at the curve's corner when the regularization norm increases immediately as $\sigma _f^2$ becomes smaller than the optimal ${\sigma _f} ^2$, which is not the case for our experiments. The values obtained at the point of maximum curvature provided better optimal estimates across all experiments, and these are the values listed under the L-curve row in \cref{tab:2}.

After obtaining the estimated variances across all $500$ runs, we observed outlier in the GCV and $\chi ^2$ estimates, which were removed based on experiment-specific thresholds: $[0.35, 0.7]$ in experiment 1, $[0.003, 0.7]$ in experiment 2, $[0.8, 10]$ in experiment 3, and $[0.5, 6]$ in experiment 4. The resulting mean and standard deviation of the estimates are are summarized in \cref{tab:2}.
\FloatBarrier
\begin{table}[tbhp]
\centering
\footnotesize
\renewcommand{\arraystretch}{1.2}
% \resizebox{\textwidth}{!}{%
\begin{tabular}{|c|c|c|c|} 
\hline
\textbf{Expt} & \textbf{L-curve} & \textbf{GCV} & \boldmath$\chi^2$ \\
\specialrule{1.2pt}{0pt}{0pt}
1 & 0.5291 (0.0773) & 0.5413 (0.0895) & 0.5332 (0.0984) \\ \hline
2 & 0.6825 (0.0866) & 0.0496 (0.0096) & 0.0044 (0.0010) \\ \hline
3 & 0.9982 (0.0646) & 5.2517 (0.8821) & 2.5881 (0.6656) \\ \hline
4 & 1.1584 (0.1008) & 3.0389 (0.2422) & 2.1468 (0.3629) \\
\specialrule{1.2pt}{0pt}{0pt}
\end{tabular}%
% }
\caption{Mean (standard deviation) of ${\sigma_f}^2$ estimates from the three regularization parameter selection methods.}
\label{tab:2}
\end{table}

\Cref{tab:2} presents the mean and standard deviation of the model error variance  ${\sigma _f} ^2$ estimated using the L-curves, GCV curves, and $\chi ^2$ method across all 500 independent runs for each experiment. 
In experiments 1 and 2, the first guess is closer to the true solution than the simulated data and in experiments 3 and 4 the simulated data is more accurate than the first guess. In experiment 1, all three methods produce similar estimates for ${\sigma _f} ^2$. However, in experiment 2, ${\sigma _f} ^2$ values vary significantly. The $\chi ^2$ method yields the smallest variance, while the L-curve method gives the highest variance. This difference reflects a decreased reliance on the model dynamics in favor of data correction with the L-curve method and more trust in the model dynamics with the $\chi^2$ method. For experiment 3, the estimated ${\sigma _f} ^2$ differ considerably, with the L-curve method producing the lowest value and and the GCV method the highest. This suggests that the L-curve method gives more weight to the model than other methods, while the GCV method strongly favors the simulated data. In experiment 4, the L-curve method produces the smallest ${\sigma _f} ^2$, while the GCV method yields the highest. The $\chi^2$ method produces an intermediate value. This again indicates that the L-curve method assigns more weight to the model dynamics, while GCV and $\chi^2$ place greater emphasis on data correction, though to different extents.

Across all experiments, $\sigma_f^2$ estimates are generally lower in experiments 1 and 2 compared to experiments 3 and 4. This suggests that when the model is more accurate, the regularization methods correctly assign a lower model error variance, thereby correctly identifying that we should rely more on model. Conversely, when the observational data is more reliable, higher $\sigma_f^2$ values correctly indicate we should rely more heavily on the data. However, when $\sigma_f^2$ becomes too large, particularly with the GCV method in experiment 3, there is a problem with over-reliance on data. There are $49$ data and $89000$ state estimates, so the problem is very under-determined. If $\sigma_f^2$ is too large, the problem is ill-posed and we will see in \cref{sec:5.3} that this results in unrealistic oscillations in the state estimates.

\subsubsection{Optimal PM$_{2.5}$ estimates}\label{sec:5.3}

After obtaining $\sigma _f^2$ in for each experiment using the three regularization parameter selection methods, we use the model error variances that correspond to the first column of each dataset to do data assimilation in order to obtain PM$_{2.5}$ concentration estimates as shown in the results below.

\begin{figure}[H]
    \centering
    \begin{subfigure}[b]{0.49\textwidth}
        \includegraphics[width=\textwidth]{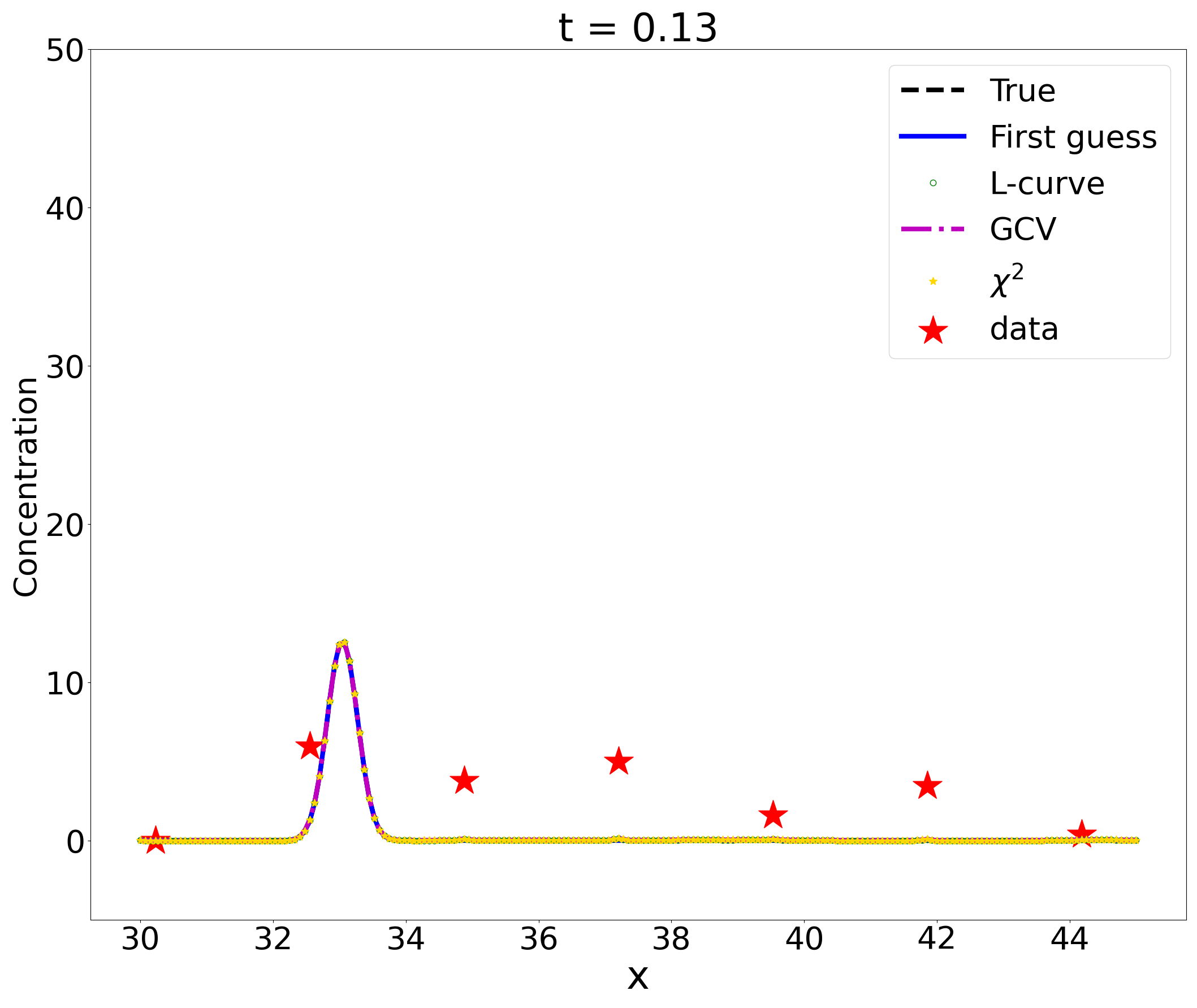}
    \end{subfigure}
    \begin{subfigure}[b]{0.49\textwidth}
        \includegraphics[width=\textwidth]{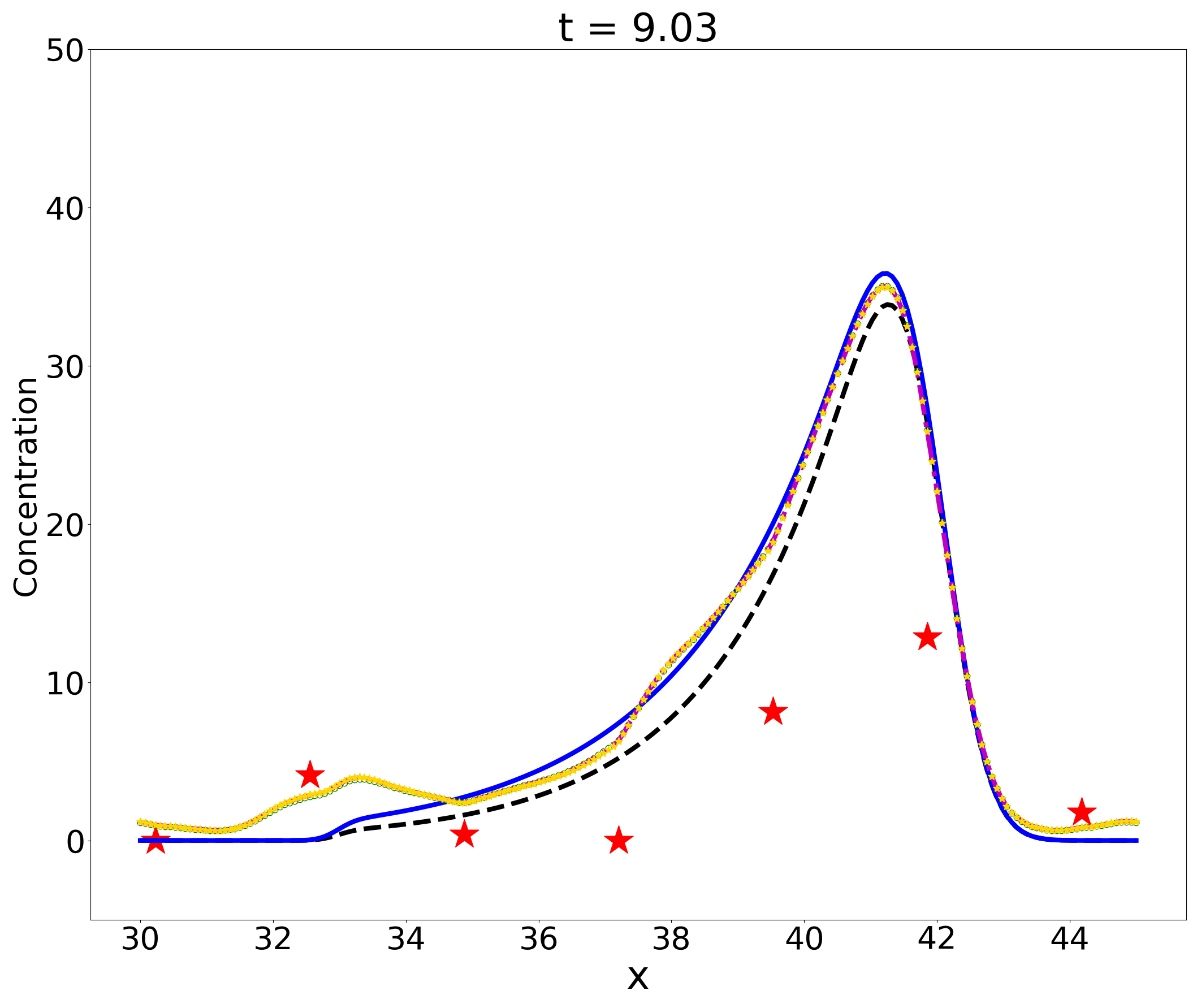}
    \end{subfigure}
    \begin{subfigure}[b]{0.49\textwidth}
        \includegraphics[width=\textwidth]{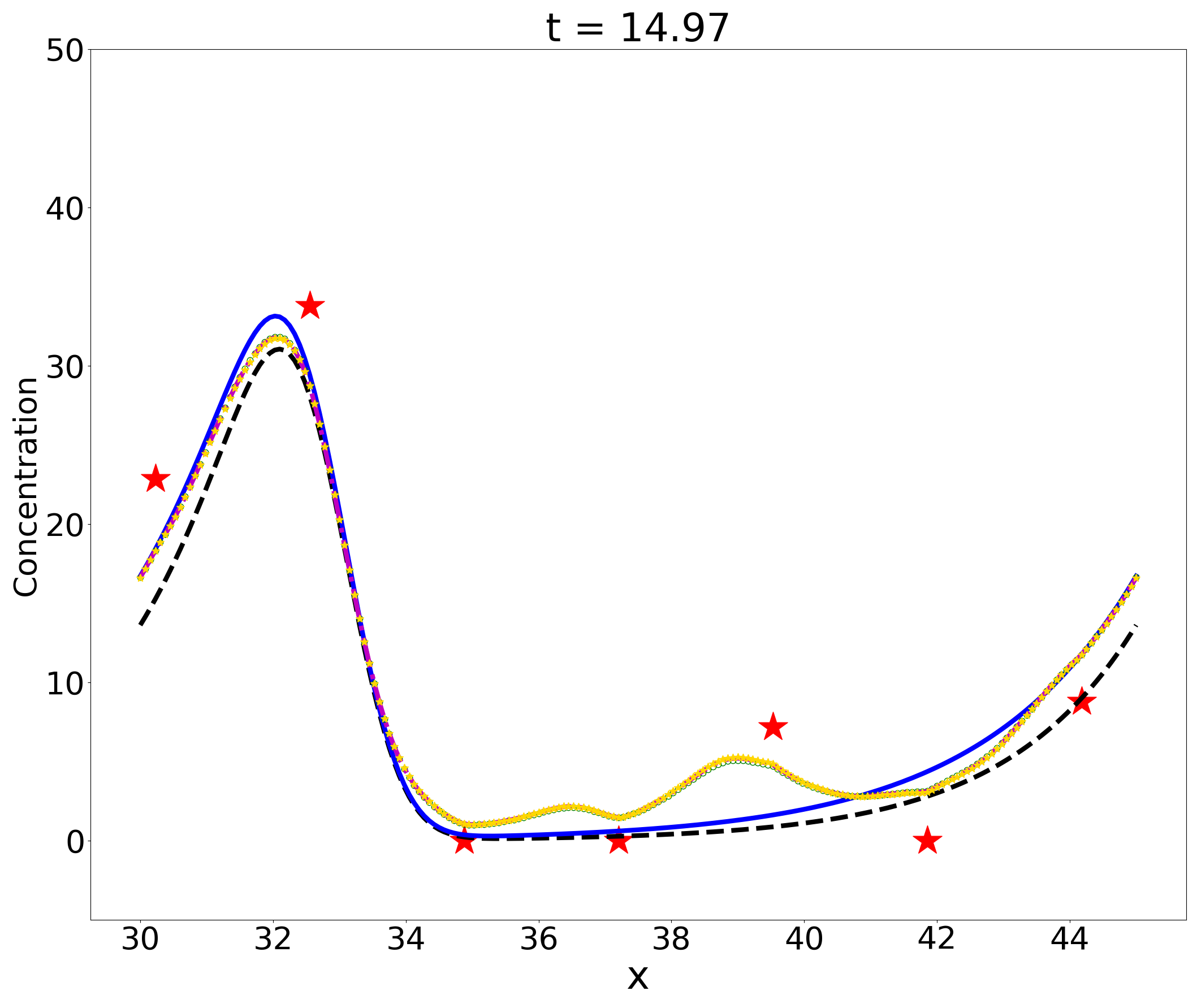}
    \end{subfigure}
    \begin{subfigure}[b]{0.49\textwidth}
        \includegraphics[width=\textwidth]{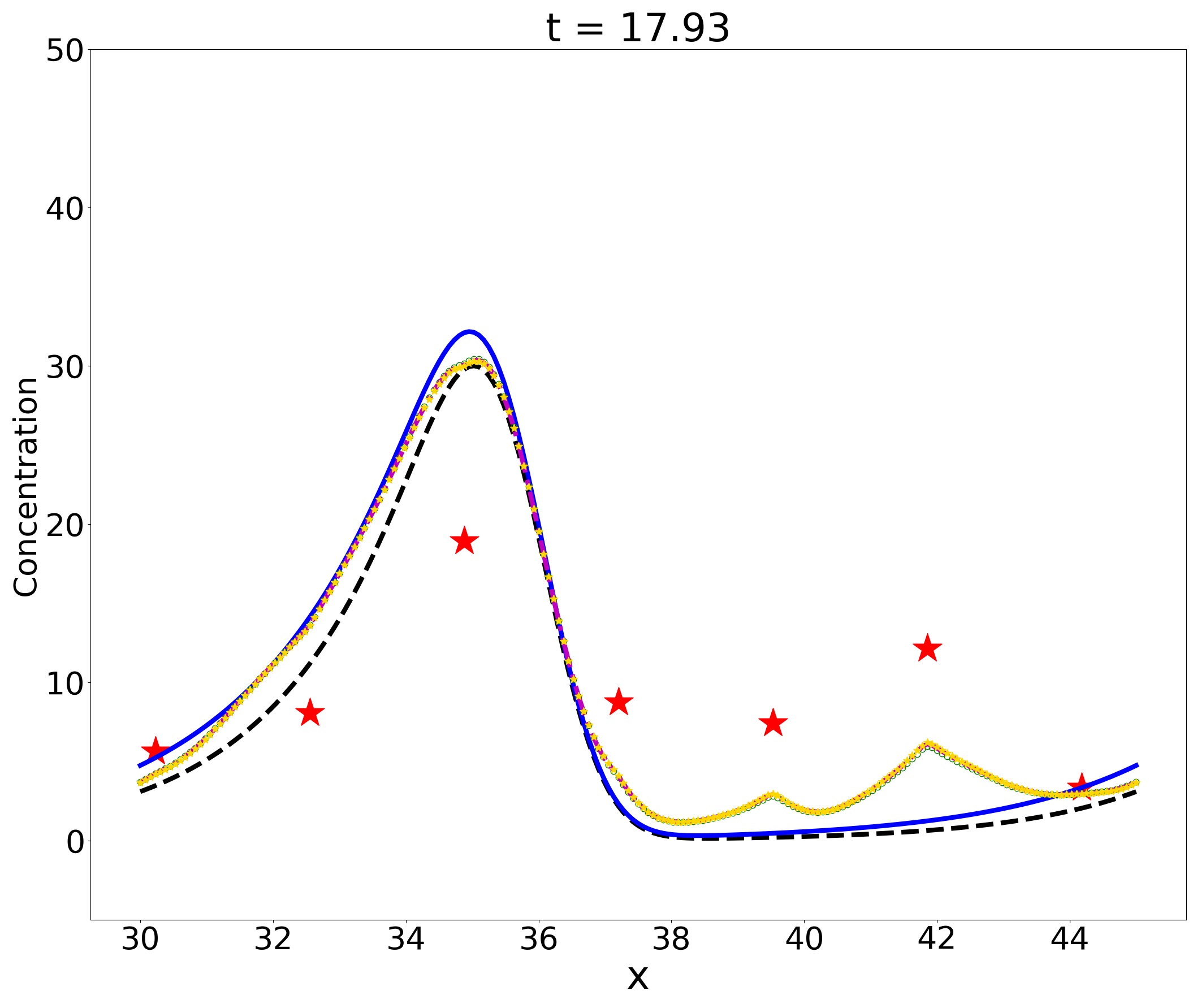}
    \end{subfigure}
    \caption{PM$_{2.5}$ concentration as a function of space for experiment 1}
    \label{fig:7}
\end{figure}

\textbf{Experiment 1.} This experiment was set up with the first guess closely approximating the true PM$_{2.5}$ concentration, more so than the simulated observational data. \Cref{fig:7} shows the spatial PM$_{2.5}$ concentrations at four distinct times during the transport process. This experiment features a single wildfire emission source located at $x=33$ and uses periodic boundary conditions. The standard deviation of the observational noise ranges from 0 to 21, while the estimated standard deviation of the first guess error is approximately 0.7.

Across all time steps, the assimilated estimates generally follow the first guess more closely than the data, consistent with the experimental design. At early times (e.g. at$t = 0.13$), the assimilated estimates align nearly exactly with both the first guess and the true solution, reflecting minimal influence from the observational data. As the transport progresses (e.g. at $t= 9.03, 14.97, 17.93$), the assimilated estimates attempt to reconcile with the observational data but still largely preserve the structure of the first guess. In some regions like at around ($t =9.03$, $x = 33$), the assimilation appear to significantly to deviate from the first guess which is more reliable, resulting in poor estimates. 

This apparent deviation arises because the optimization process, though global, cannot entirely disregard data even when those observations are inconsistent with the underlying dynamics. While this may be viewed as a failed assimilation, the proposed methods did accurately capture that the first guess is overall closer to the true solution than the data. This overall conclusion could give the user confidence to increase the weight on the first guess so that the assimilated estimate is less to fit apt the data.
\FloatBarrier

\begin{figure}[H]
    \centering
    \begin{subfigure}[b]{0.49\textwidth}
        \includegraphics[width=\textwidth]{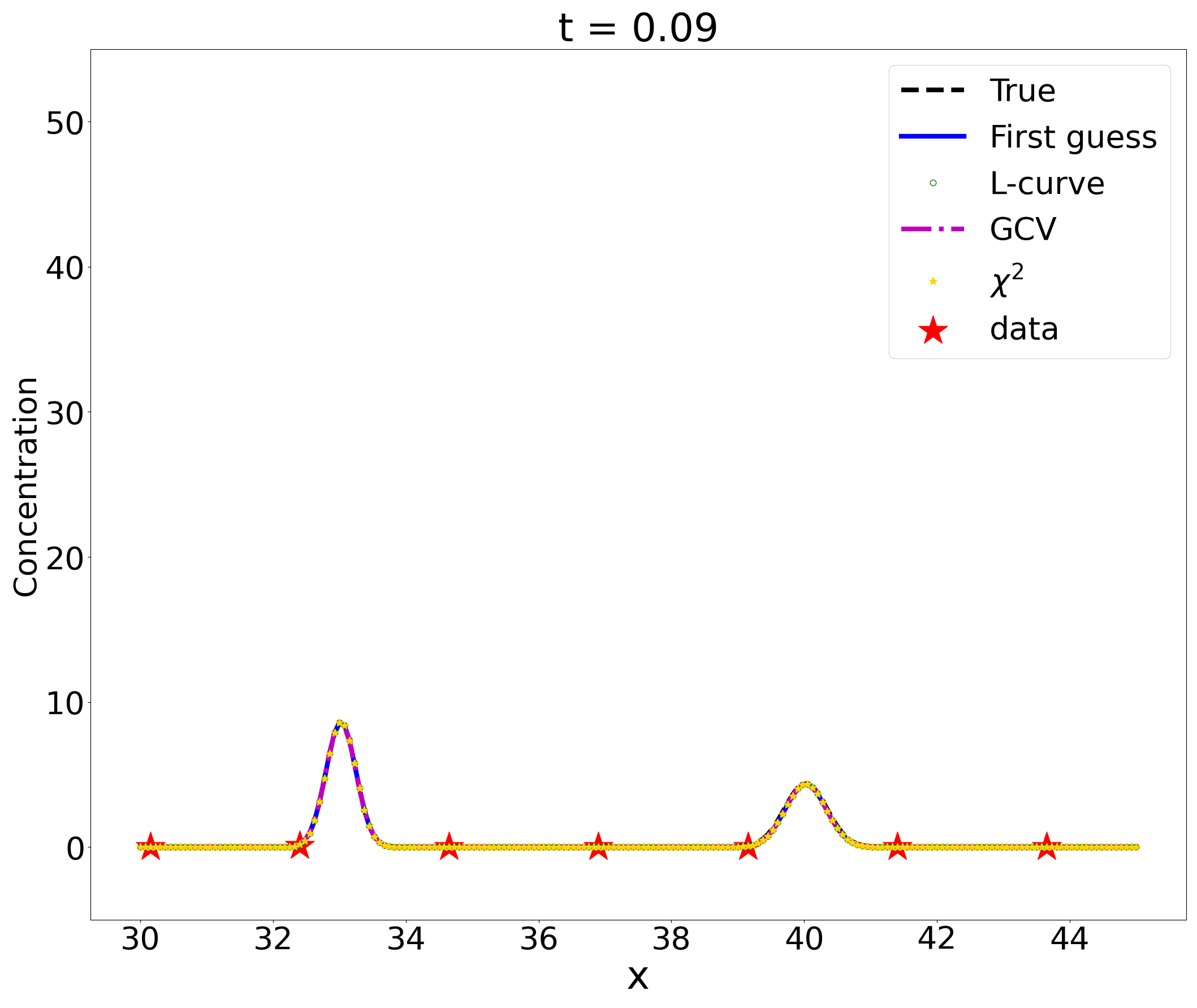}
    \end{subfigure}
    \begin{subfigure}[b]{0.49\textwidth}
        \includegraphics[width=\textwidth]{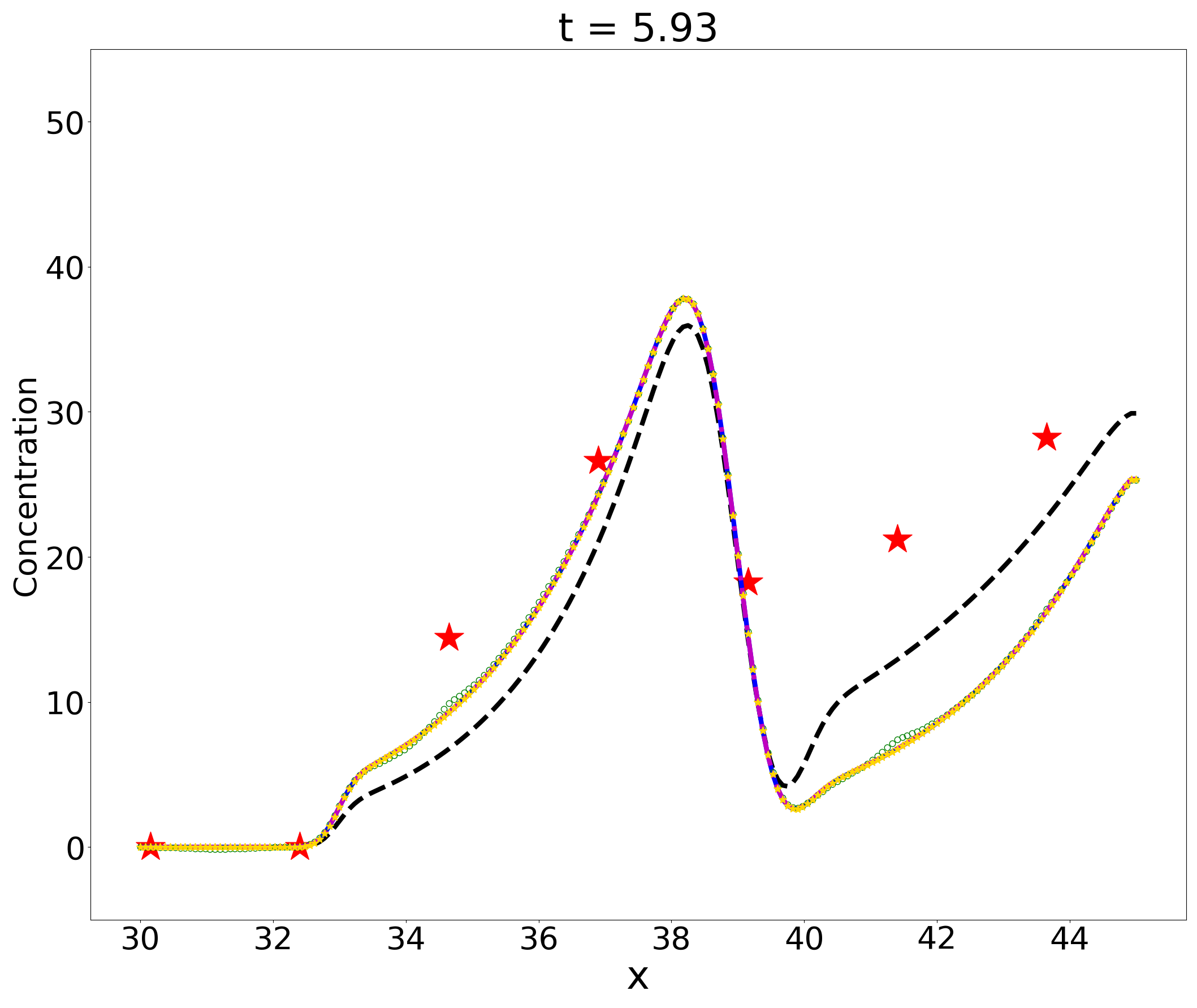}
    \end{subfigure}
    \begin{subfigure}[b]{0.49\textwidth}
        \includegraphics[width=\textwidth]{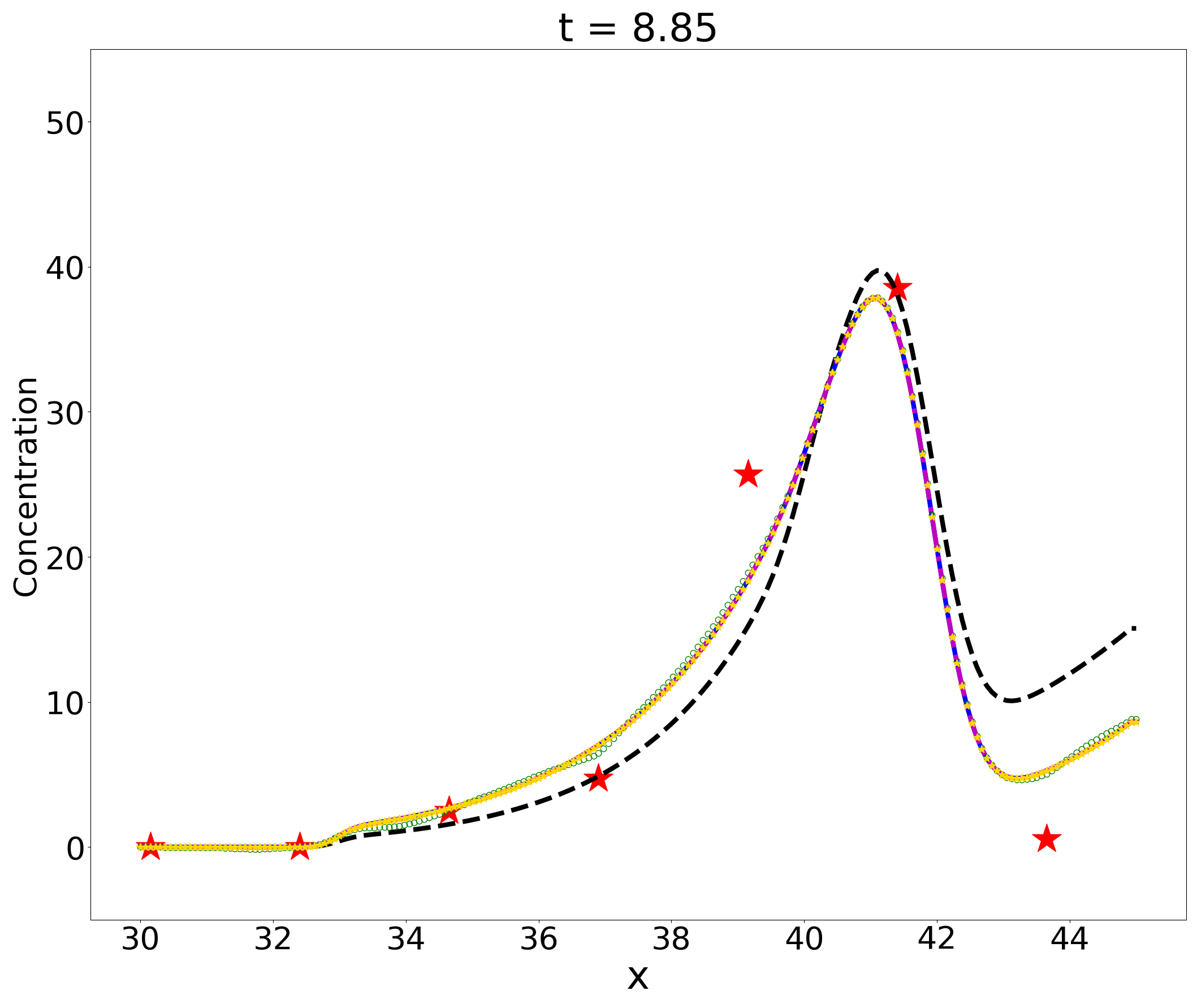}
    \end{subfigure}
    \begin{subfigure}[b]{0.49\textwidth}
        \includegraphics[width=\textwidth]{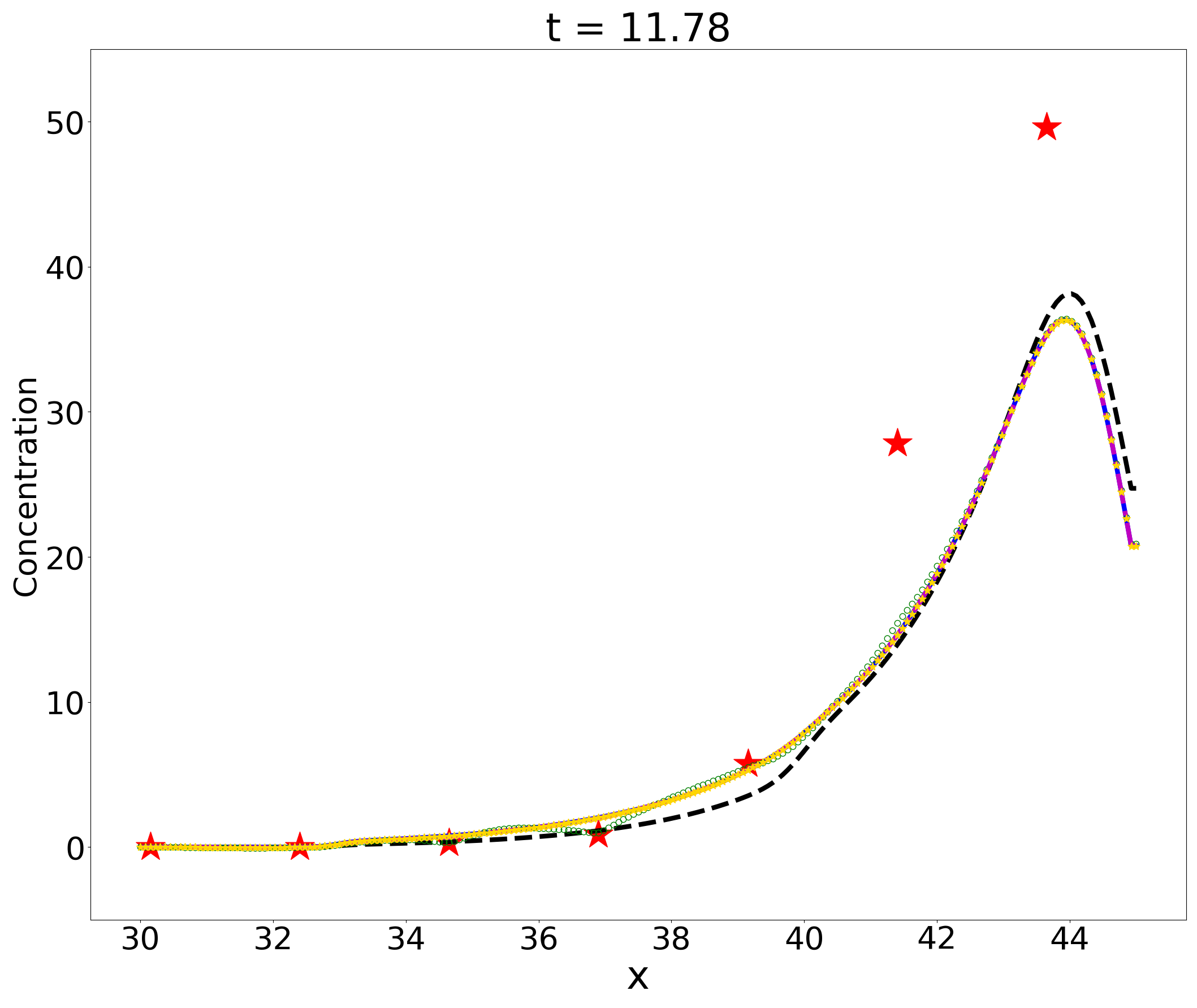}
    \end{subfigure}
    \caption{PM$_{2.5}$ concentration estimates a function of space for \textit{experiment 2}}
    \label{fig:9}
\end{figure}

\textbf{Experiment 2.}
This experiment was set up similarly to experiment 1, but the difference in this experiment is that it has two sources of PM$_{2.5}$ at $x=33$ and $x=40$, and no flux boundary conditions were used. \Cref{fig:9} presents the spatial PM$_{2.5}$ concentration at four points in time during the transport process. 

Across all times, the assimilated estimates from all three methods remain closely aligned with the first guess. This behavior is largely attributed to the relatively high uncertainty in the observational data, with noise standard deviations ranging from 3.84 to 6.72, compared to significantly smaller estimated model error variances (on the order of 0.8 or lower). At the initial time ($t = 0.13$), the assimilated state estimates, observational data, first guess, and true concentration values all align closely across the spatial domain. As the transport process evolves ($t = 05.93$ and $t = 8.85$),  discrepancies between the first guess and the true solution become more noticeable. However, the assimilation continues to prioritize the first guess, only fitting data points that align with it. At $t = 11.78$, the L-curve estimates display a clear effort to better fit the third, fourth and fifth data points, showing a significant deviation from the first guess. This deviation is an indication of the L-curve choice of larger variance in the model. Assimilated estimates from the GCV and $\chi ^2$ methods, on the other hand remain closely aligned with the first guess, fitting only the observed data that align with the first guess.

Compared to Experiment 1, the results here demonstrate more stable assimilation behavior, possibly due to the use of no-flux boundary conditions rather than the periodic boundary condition used in Experiment 1. Overall, this experiment underscores the importance of variance estimation in guiding the assimilation process and justifying reliance on the model when data uncertainty is high.

\FloatBarrier

\begin{figure}[H]
    \centering
    \begin{subfigure}[b]{0.49\textwidth}
        \includegraphics[width=\textwidth]{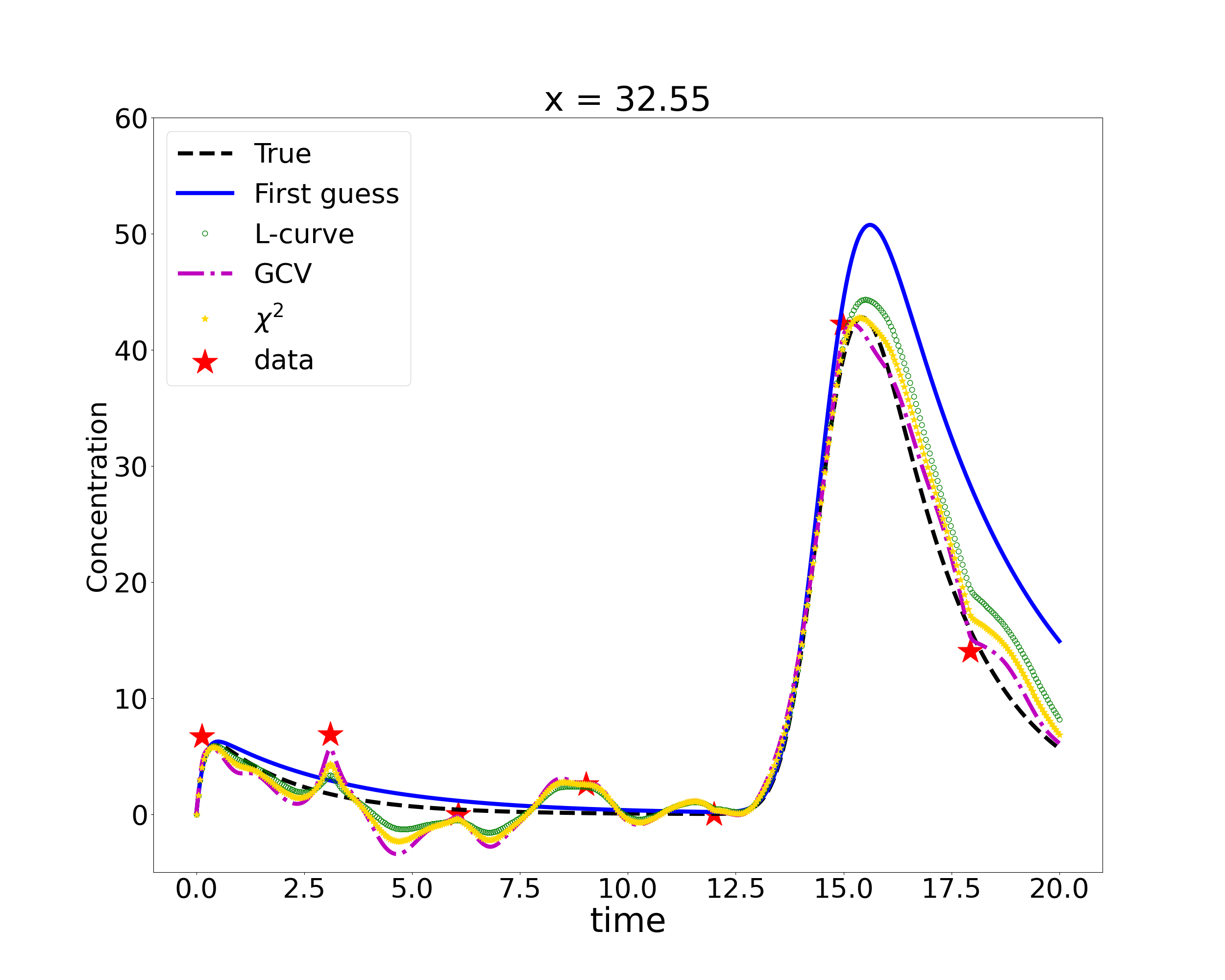}
    \end{subfigure}
    \begin{subfigure}[b]{0.49\textwidth}
        \includegraphics[width=\textwidth]{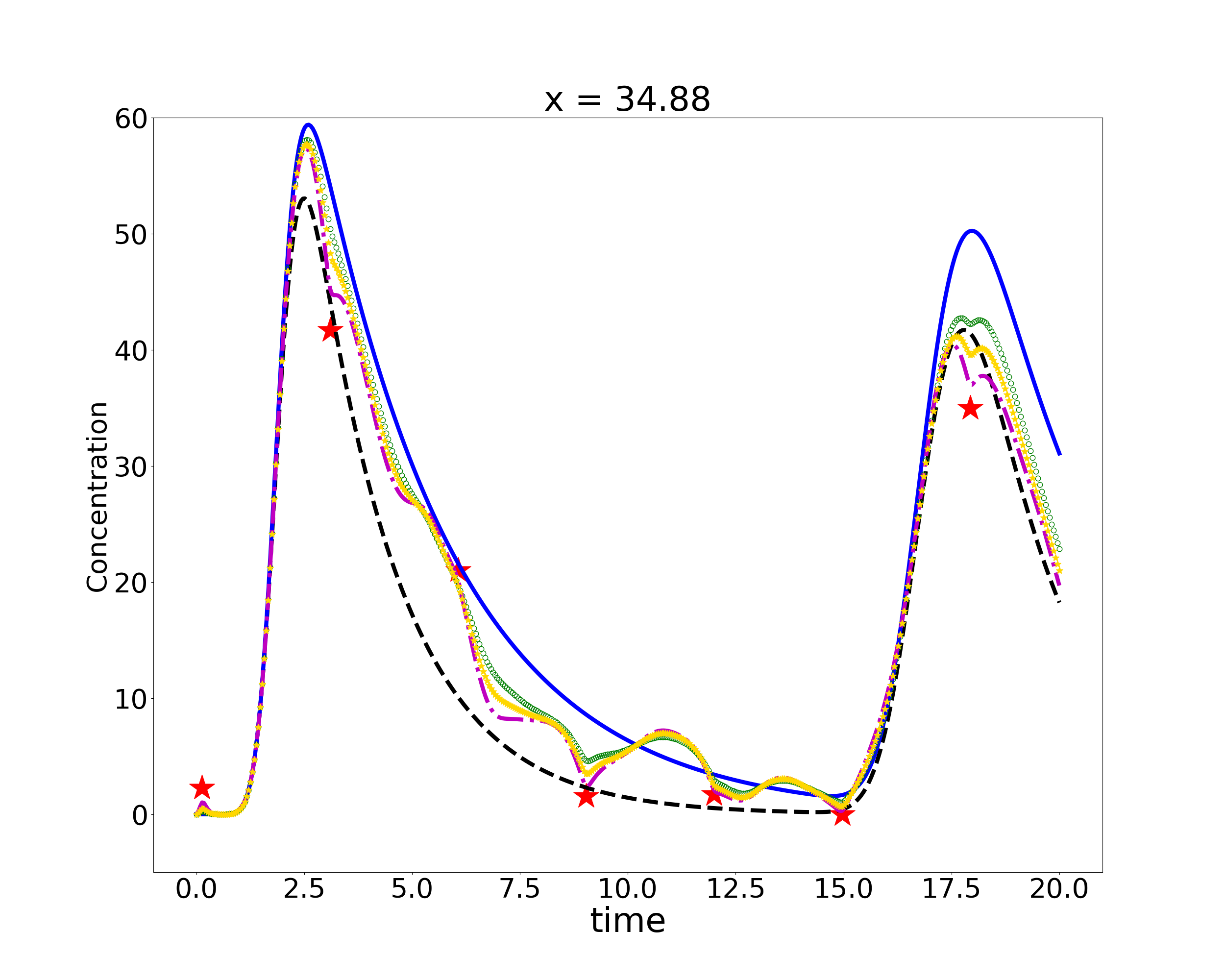}
    \end{subfigure}
    
    \begin{subfigure}[b]{0.49\textwidth}
        \includegraphics[width=\textwidth]{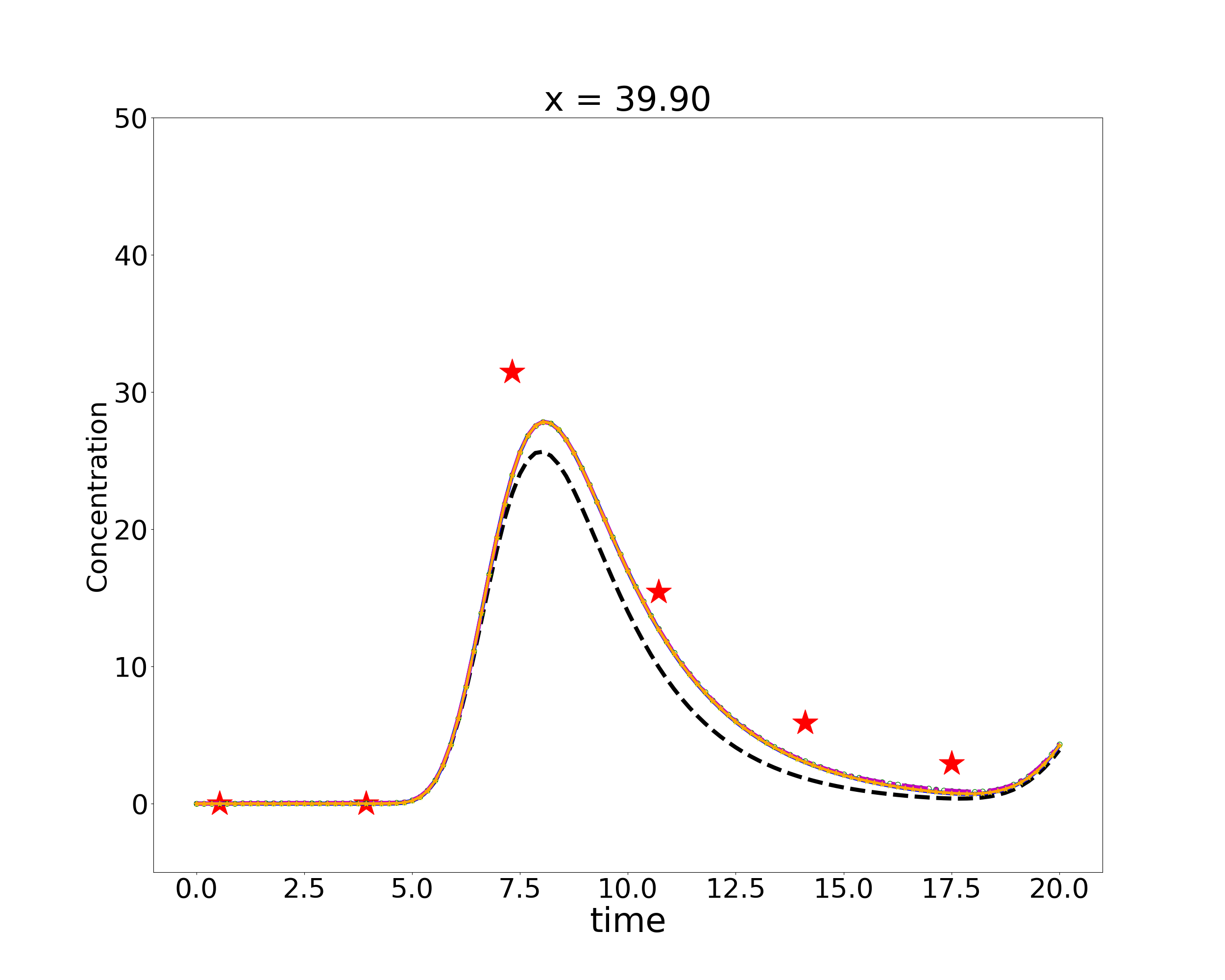}
    \end{subfigure}
    \begin{subfigure}[b]{0.49\textwidth}
        \includegraphics[width=\textwidth]{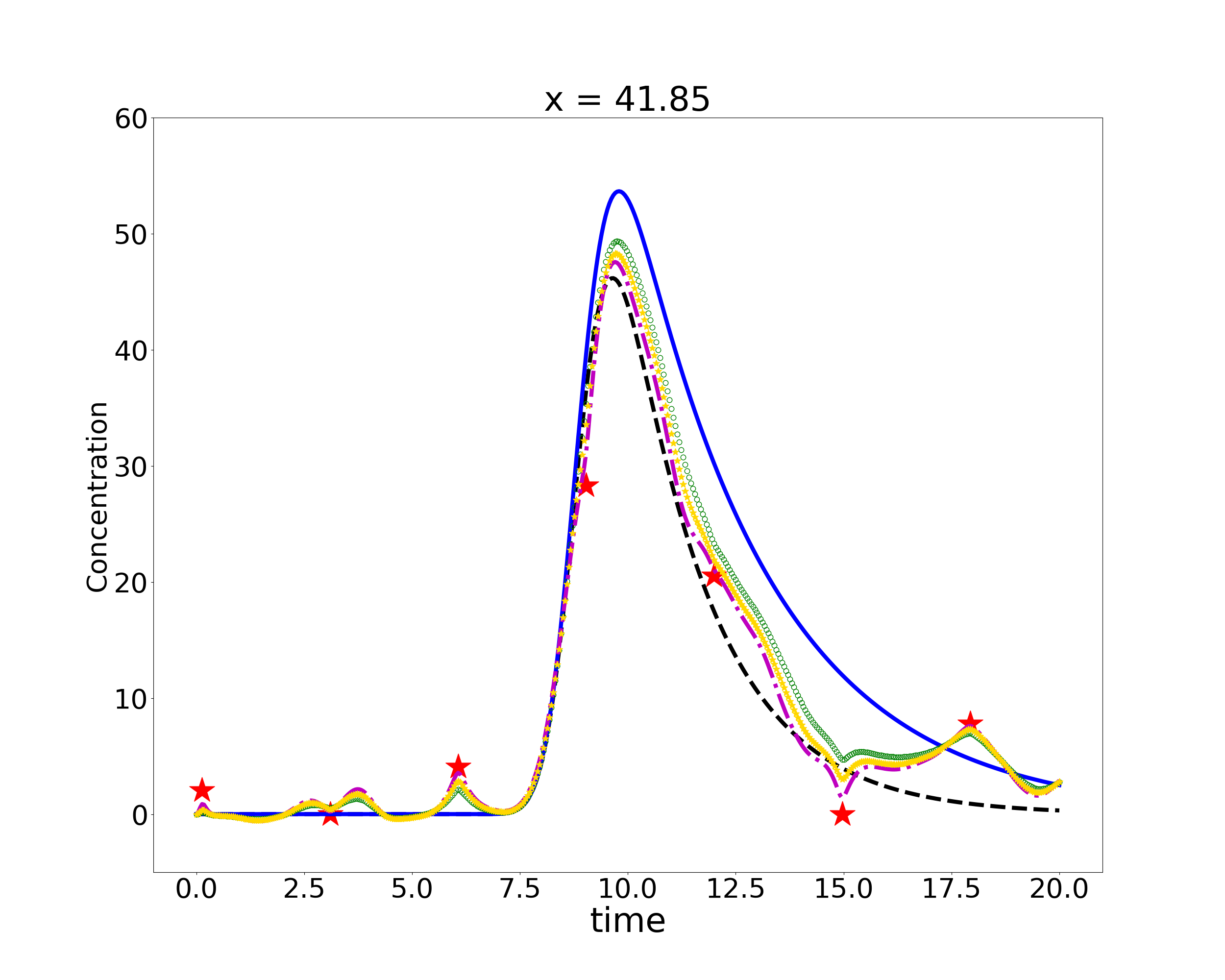}
    \end{subfigure}
    
    \caption{ PM$_{2.5}$ concentration estimates a function of time for \textit{experiment 3}}
    \label{fig:4}
\end{figure}

\textbf{Experiment 3.}
This experiment was designed with the simulated observational data more accurate than the first guess. Similar to experiment 1 a single source is located at $x=33$ and periodic boundary conditions were used.  \Cref{fig:4} shows PM$_{2.5}$ concentration as a function of time at four spatial locations. These locations were strategically chosen to capture critical moments in the transport process, focusing on positions before and after the source of the pollutant.

Across all locations, the assimilated estimates show a clear shift toward fitting the observational data, particularly for the GCV method, which reflects its higher estimated model error variance. At $x=32.55$, all assimilated estimates attempt to fit the data points but these estimates exhibit noticeable oscillations between $t=5$ and $t=15$, likely due to ill-posedness in the underdetermined problem. These osccilation are also noticeable at the other time locations. At $x=34.88$, the assimilated estimates try to fit the data but at the impulse at $t=2.5$, they more closely align with the first guess. After this point, the estimates from different methods slightly differ. The estimates from the GCV method fit the data points more closely. All three assimilated estimates deviate significantly from the first guess in an attempt to fit the data. At $x=37.20$ and $x=41.85$, all three assimilated estimates attempt to fit the data, showing a stronger preference to the observational data than the first guess. The GCV estimates fit the data points more closely  than other methods reflecting the larger value of $\sigma _f^2$ that was chosen with this method.

\FloatBarrier

\begin{figure}[H]
    \centering
    \begin{subfigure}[b]{0.49\textwidth}
        \includegraphics[width=\textwidth]{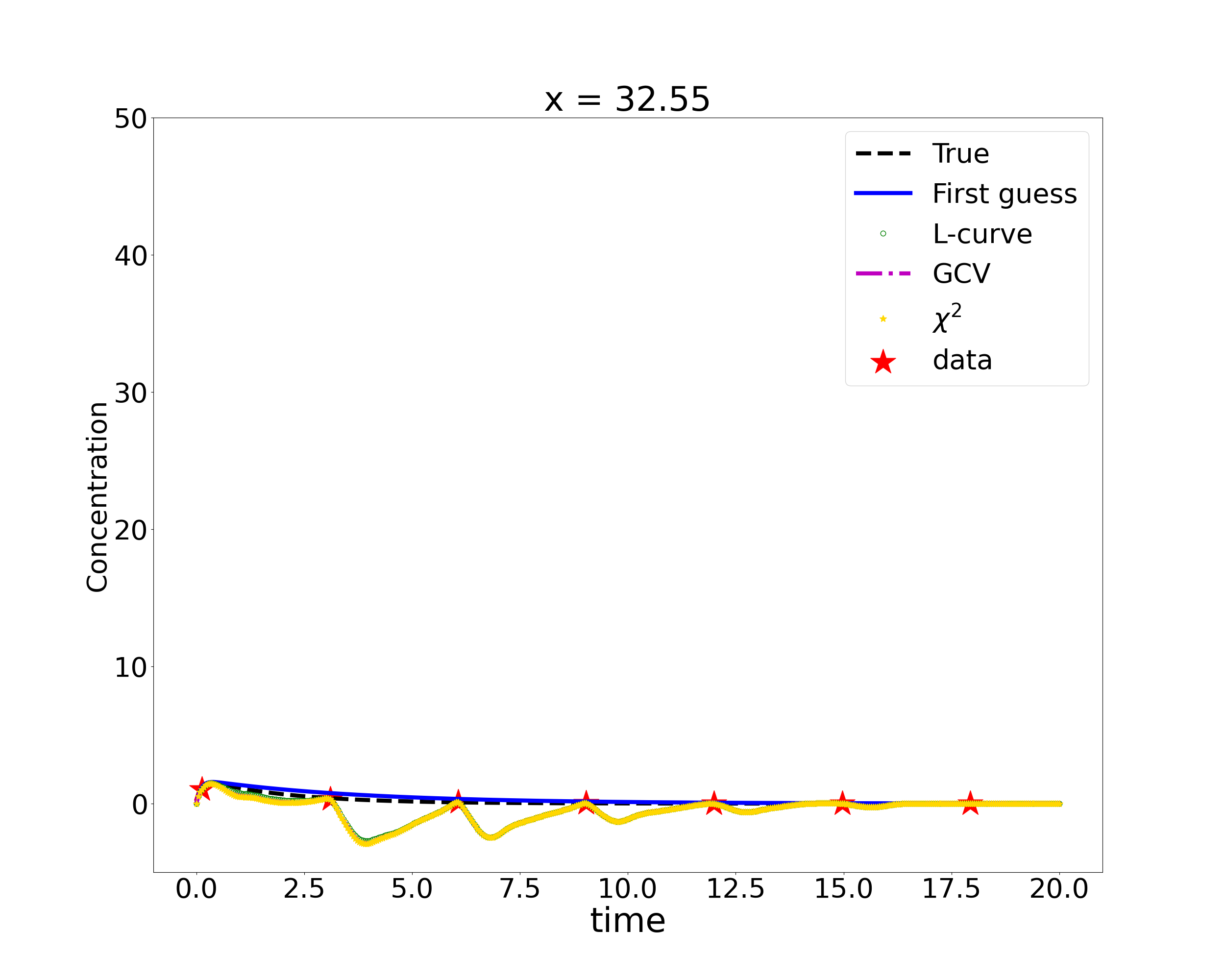}
    \end{subfigure}
    \begin{subfigure}[b]{0.49\textwidth}
        \includegraphics[width=\textwidth]{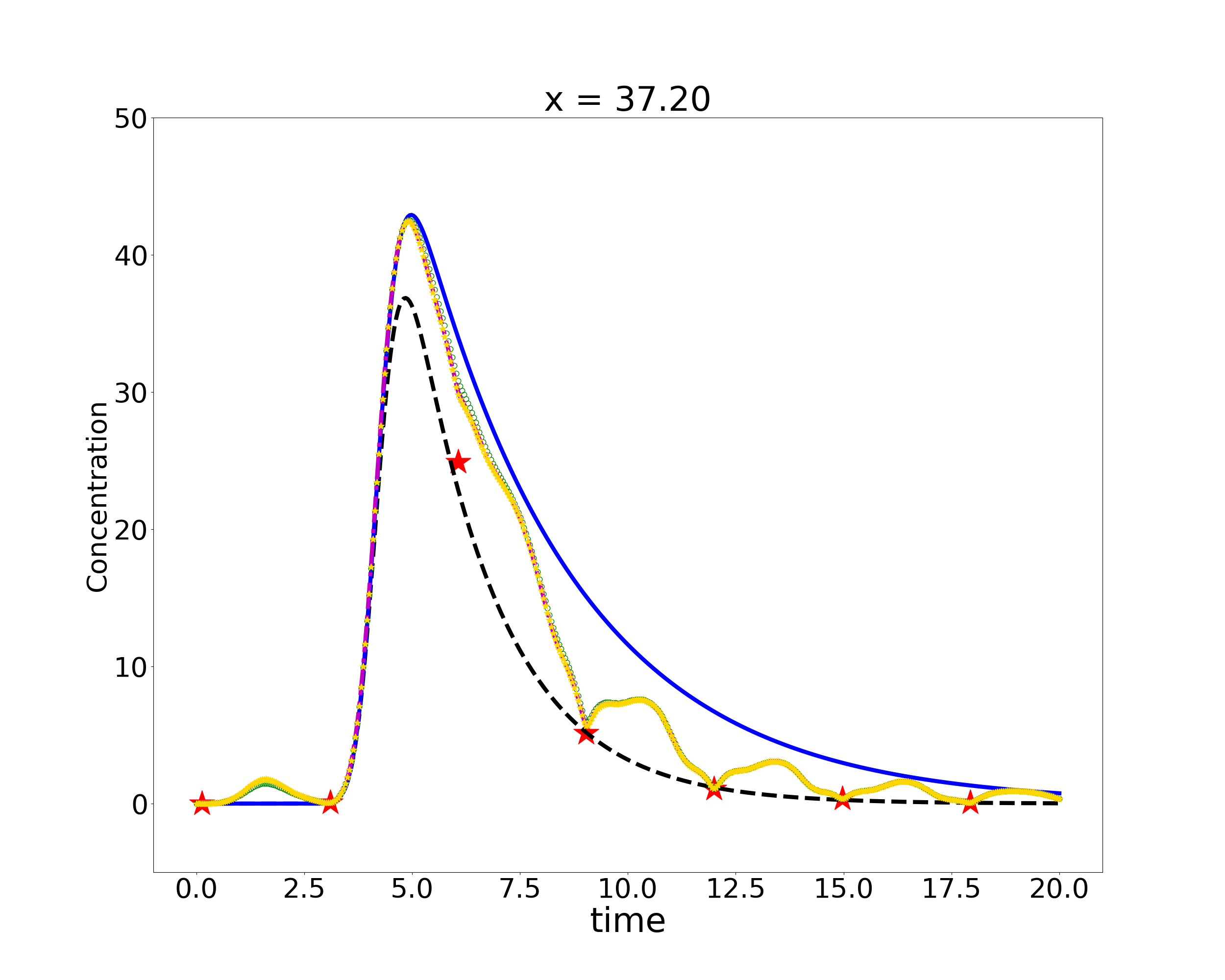}
    \end{subfigure}
    \begin{subfigure}[b]{0.49\textwidth}
        \includegraphics[width=\textwidth]{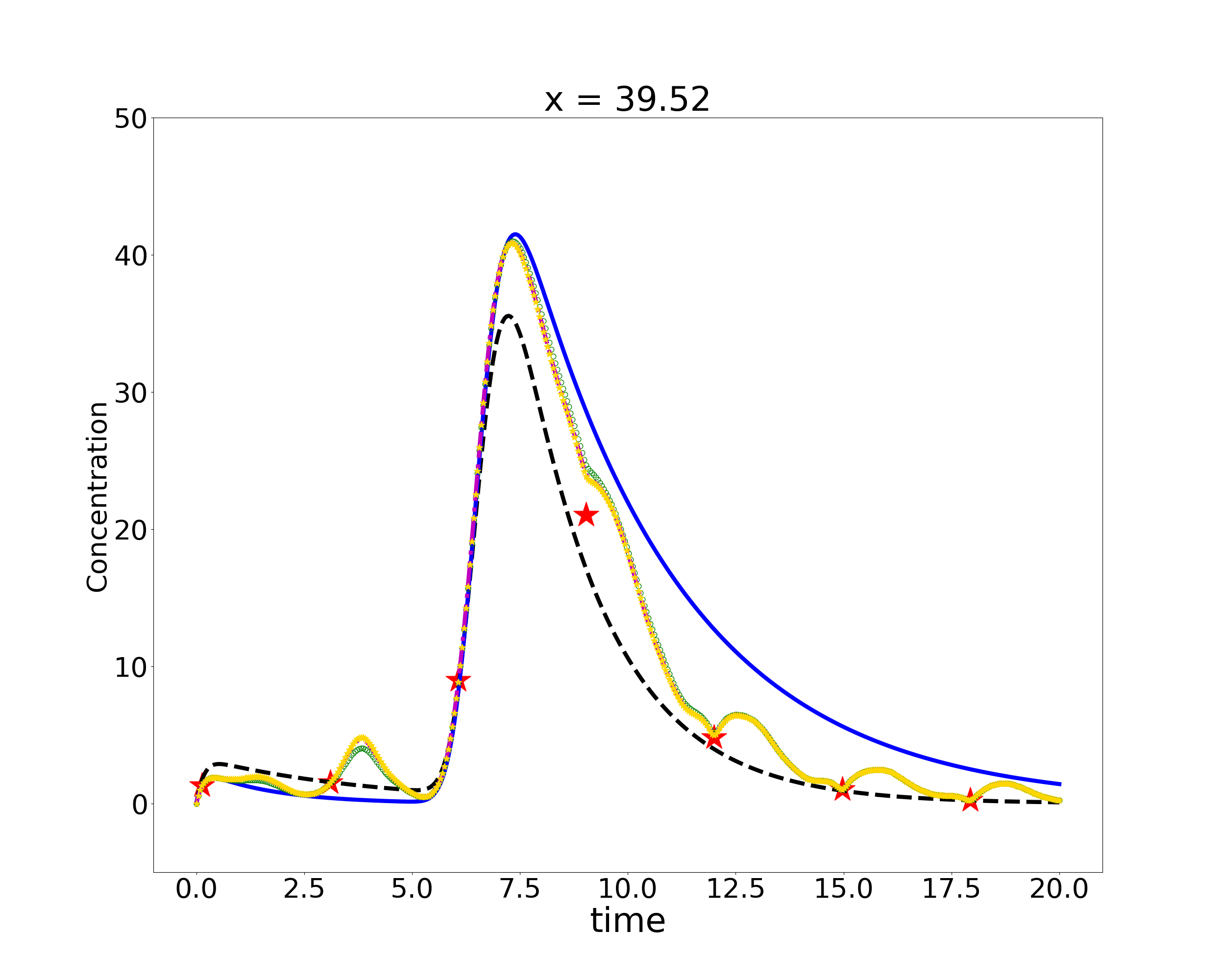}
    \end{subfigure}
    \begin{subfigure}[b]{0.49\textwidth}
        \includegraphics[width=\textwidth]{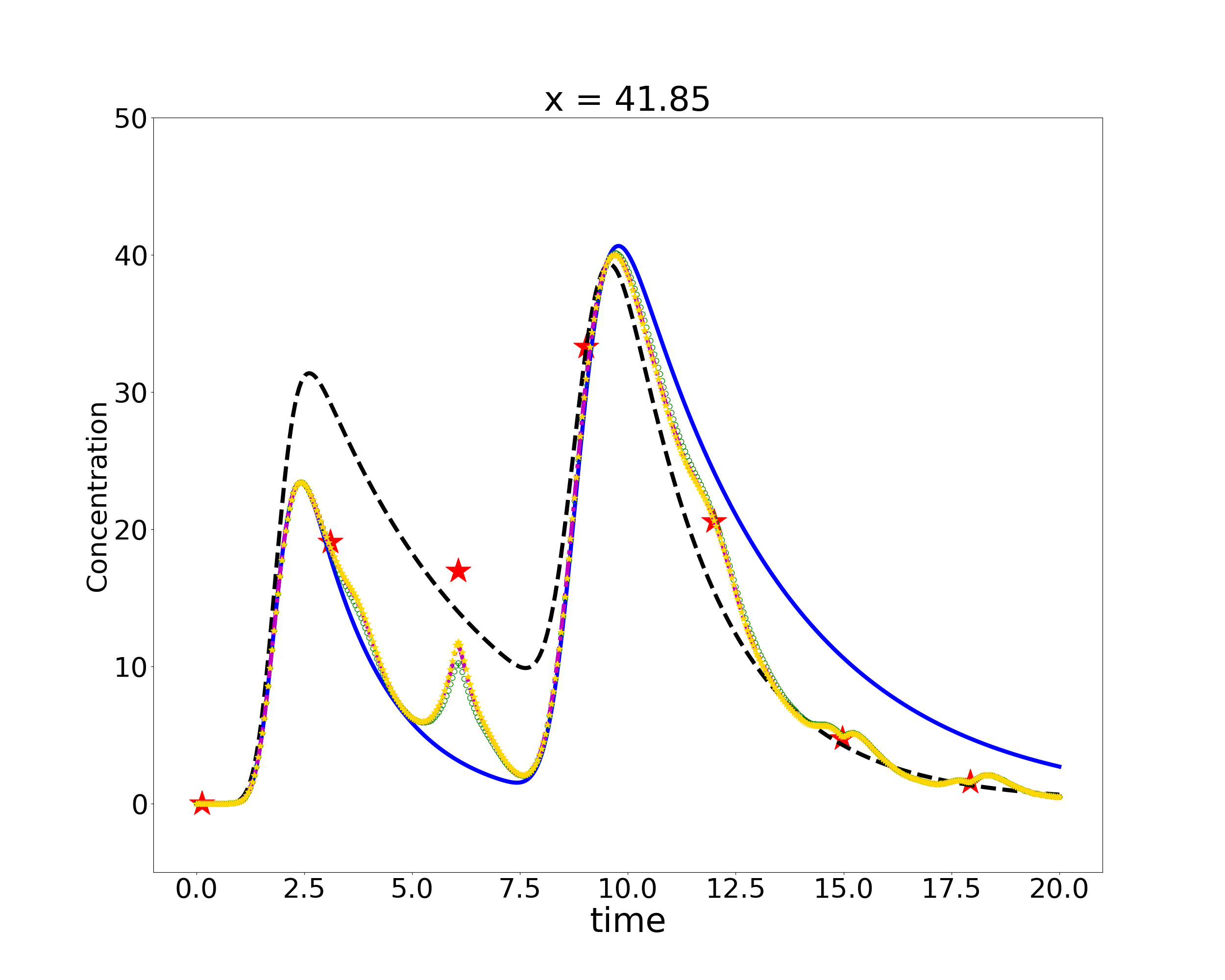}
    \end{subfigure}
    \caption{PM$_{2.5}$ concentration estimates a function of time for \textit{experiment 4}}
    \label{fig:5}
\end{figure}

\textbf{Experiment 4} 
This experiment was set up similarly to experiment 3, but the key difference in this experiment is that it has two sources of PM$_{2.5}$ located at $x=33$ and $x=40$, and no flux boundary conditions were used. \Cref{fig:5} presents the temporal evolution of PM$_{2.5}$ concentration estimates at four spatial locations.

At $x = 32.55$, a location on the left side of both emission sources, the PM$_{2.5}$ concentration is close to zero due to the constant wind field used, blowing the pollutant to the right. The assimilated estimates oscillate as they fit the data. This oscillation can be attributed to the ill-posed nature of the underdetermined problem. At $x=37.2$ and $x=39.52$, the assimilated estimates show a clear attempt to fit the data points but they also follow the first guess while making necessary adjustments to accommodate the observed data. This balance between following the first guess and adapting to the data points illustrates the estimates' capacity to capture the key features of the transport process at these locations. At $x=41.85$,  the assimilated estimates manage to largely follow the first guess while making necessary adjustments to accommodate the observed data at most times except for the attempt to fit data near $t=7.5$. This balance between following the first guess and adapting to the data points illustrates the estimates' capacity to capture the key features of the transport process at this location.

\FloatBarrier

\subsubsection{Root mean square errors (RMSE)}
\Cref{tab:3} presents the RMSE of the three PM$_{2.5}$ concentration estimates and the true concentrations. 
\FloatBarrier
\begin{table}[tbhp]
\centering
\footnotesize
\renewcommand{\arraystretch}{1.2}
% \resizebox{\textwidth}{!}{%
\begin{tabular}{|c|c|c|c|c|c|} 
\hline
\textbf{Expt} & \textbf{First guess} & \textbf{Data} & \textbf{L-curve} & \textbf{GCV} & \boldmath$\chi^2$ \\ 
\specialrule{1.2pt}{0pt}{0pt}
1 & 1.5319 & 5.3388 (0.3745) & 1.6997 (0.2050) & 1.7146 (0.2625) & 1.8155 (0.3255) \\ \hline
2 & 2.1465 & 5.6288 (0.7138) & 2.0477 (0.0580) & 2.1220 (0.0108) & 2.1405 (0.0021) \\ \hline
3 & 6.5241 & 3.2949 (0.2154) & 3.6945 (0.2779) & 2.9229 (0.2668) & 3.1137 (0.3822) \\ \hline
4 & 5.8753 & 2.2268 (0.3068) & 3.9227 (0.0961) & 3.5954 (0.1446) & 3.7035 (0.1714) \\  \specialrule{1.2pt}{0pt}{0pt}
\end{tabular}%
 % }
\caption{RMSE between the assimilated PM$_{2.5}$ estimates and the true concentration for each experiment. Reported values represent the mean RMSE, with standard deviations shown in parentheses.}
\label{tab:3}
\end{table}
In experiments 1 and 2, the first guess RMSE is significantly lower than the data mean RMSE indicating that the model is initially more accurate than the observations. In experiment 1, the mean RMSEs of the assimilated estimates (L-curve, GCV, and $\chi ^2$) are slightly higher than the first guess but remain substantially lower than the data error. This suggests that the automated selection of $\sigma _f^2$ correctly identified the model's higher accuracy and maintained its influence in the assimilation process while limiting the effect of noisy data. In experiment 2, the first guess and assimilated estimates have similar or lower mean RMSEs. This indicates that the assimilation process preserves the reliability of the model while limiting influence from noisy observations.

In experiments 3 and 4, where the data are more accurate, the assimilated mean RMSEs fall between those of the first guess and the data. Notably, in experiment 3, the GCV method achieves the lowest mean RMSE below  that of the data itself, suggesting that it assigns a very large $\sigma_f^2$, effectively down-weighting the model and relying almost entirely on the data. While this leads to a reduced mean RMSE, it raises concerns about ill-posedness due to over-reliance on sparse observations. Given that the problem is highly under-determined (49 observations vs. 89,000 state estimates), an excessively large $\sigma_f^2$ increases sensitivity to noise, potentially leading to spurious oscillations in the assimilated estimates. Thus, while the lower mean RMSE from GCV might suggest an improvement in accuracy, the high variance estimate implies a loss of regularization, making the solution more prone to instability. 

Across all experiments, the regularization-based assimilation methods (L-curve, GCV, and $\chi^2$) consistently improve upon the worst-performing input (either first guess or data). The mean RMSE of the assimilated estimates remain lower than the largest error in each case, confirming that the regularization parameter selection methods effectively balance model fidelity and observational accuracy.

\subsection{ Non-isotropic case}
\subsubsection{Estimation of model error covariance hyperparameters}

We implemented  the  non-isotropic model error covariance $C_f(\mathbf{x},t,\mathbf{x}',t')$ introduced in section \ref{subsec_3.2.2}, which incorporates both spatial and temporal correlations.  This results in a full matrix and so we reduced the grid resolution to 50 spatial points and 112 time steps, resulting in a total state size of $N=5763$. We estimate three hyperparameters, i.e., the model error variance $\sigma_f^2$, spatial correlation length $l_f$, and temporal correlation scale $\tau_f$ using both the GCV and $\chi^2$ methods. These hyperparameters were optimized under bounded domains: $\sigma_f^2 \in [10^{-6},9]$, $l_f \in [1,15]$ and $\tau_f \in [1,20]$. These bounds were chosen based on the spatial domain $x \in [30,45]$ and the temporal interval $t \in [0,25]$, ensuring that the estimated correlation scales are meaningful relative to the extent of smoke transport in both space and time. Bounding $l_f$ and $\tau_f$ ensures that the covariance structure captures realistic plume behavior correlations that are neither overly local (leading to noise amplification) nor overly global (resulting in artificial smoothing). Similarly, bounding $\sigma_f^2$ limits the strength of the model error to plausible levels, preventing instability or overfitting in the assimilation process. 
\FloatBarrier
\begin{table}[tbhp]
\centering
\footnotesize
\renewcommand{\arraystretch}{1.2}
% \resizebox{\textwidth}{!}{%
\begin{tabular}{!{\vrule width 1.2pt}c!{\vrule width 1.2pt}c|c|c!{\vrule width 1.2pt}c|c|c!{\vrule width 1.2pt}} 
\hline
\textbf{Experiment} 
& \multicolumn{3}{c!{\vrule width 1.2pt}}{\textbf{GCV}} 
& \multicolumn{3}{c!{\vrule width 1.2pt}}{\boldmath$\chi^2$} \\ 
\cline{2-7}
& $\sigma_f^2$ & $l_f$ & $\tau_f$ & $\sigma_f^2$ & $l_f$ & $\tau_f$ \\ 
\specialrule{1.2pt}{0pt}{0pt}
1 &  0.000556 & 3.000118 & 5.000048 & 0.000194 & 3.000528 & 5.000171 \\
\hline
2 & 0.000015 &  3.007716 & 5.000869 & 0.000001 & 4.517653 & 4.023668 \\
\hline
3 & 0.005569 & 6.002482 & 12.001055 & 0.668314 & 8.575857 & 12.248493 \\
\hline
4 & 4.457971 & 12.000000 & 1.000000 & 0.027128 & 1.000003 & 15.594117 \\
\specialrule{1.2pt}{0pt}{0pt}
\end{tabular}%
% }
\caption{$\sigma_f^2$, $l_f$ and $\tau_f$ estimates obtained using the GCV and $\chi^2$ methods.}
\label{tab:4}
\end{table}

\cref{tab:4} presents the optimal model error hyperparameters, i.e., the model error variance $\sigma_f^2$, spatial correlation length $l_f$, and temporal correlation scale $\tau_f$ estimated using both the GCV and $\chi^2$ methods for the non-isotropic case. These estimates reflect the structure of the full space-time model error covariance and provide insight into how the assimilation process balances model dynamics and observational data under spatial and temporal correlation assumptions.

Across all experiments, the trends in $\sigma_f^2$ are broadly consistent with the results from the isotropic case. Specifically, in Experiments 1 and 2, where the first guess is more accurate than the data, both GCV and $\chi^2$ method yield very small values of $\sigma_f^2$  (on the order $10 ^{-4}$ or smaller), reflecting the strong trust in the model dynamics. This agreement with the isotropic case confirms that the hyperparameter selection methods correctly identify and preserve the influence of the more reliable model when data is noisy or less informative. In contrast, in Experiments 3 and 4, where the data are more accurate than the model, both methods produce larger values of $\sigma_f^2$, indicating a shift toward observational data. This shift is especially pronounced under the GCV method in Experiment 4, where $\sigma_f^2 = 4.458$, suggesting substantial model error and hence a greater reliance on the data. This trend mirrors the isotropic case, where GCV also favored larger variance estimates in data-dominated settings. However, the $\chi^2$ method  more conservative estimates of $\sigma_f^2$, particularly in Experiment 4, where it estimates a value nearly two orders of magnitude smaller than GCV. This difference underscores the greater regularization imposed by the $\chi^2$ method in non-isotropic settings, potentially offering more stable estimates under uncertainty.

The spatial and temporal correlation scales: $l_f$ and $\tau_f$ also align well with physical intuition and the characteristics of each experimental setup. In Experiments 1 and 2, both length scales remain small (around 3 to 5), consistent with a localized emission source and limited spatial-temporal spread. These values imply that model errors are expected to decorrelate quickly, which is appropriate given the relatively simple source configuration and the dominance of the first guess. In Experiment 3, both methods give larger values for $l_f$ and $\tau_f$ especially with the $\chi^2$ method. This suggests broader spatial and temporal influence of model errors, consistent with a more dominant data signal and increased complexity in the evolving plume. In Experiment 4, the GCV method selects a long spatial scale ($l_f = 12$) and a short temporal scale ($\tau_f = 1$), while the 
$\chi^2$ method selects the opposite i.e., a short spatial scale ($l_f = 1$) and a long temporal scale ($\tau_f =15.594$). This divergence reflects differences in how each method resolves the interaction between two emission sources and the temporal structure of the observed signal. The GCV method appears to assume that spatial correlations dominate, possibly over-smoothing, while the 
$\chi^2$ method favors strong temporal coherence, possibly better preserving dynamic features. 

Overall, these results demonstrate that the non-isotropic formulation captures additional complexity in the model error structure beyond what is possible in the scalar isotropic setting. Both the GCV and $\chi^2$ methods respond appropriately to whether the data or first guess is more reliable, but they differ in how they balance spatial and temporal correlations highlighting the flexibility and interpretability of the non-isotropic approach.

\subsubsection{Optimal PM$_{2.5}$ estimates}
Using the estimated values of $\sigma_f^2$, $l_f$, and $\tau_f$ from \cref{tab:4}, we perform weak-constraint 4D-Var data assimilation to obtain optimal PM$_{2.5}$ concentration estimates.

\begin{figure}[H]
    \centering
    \begin{subfigure}[b]{0.49\textwidth}
        \includegraphics[width=\textwidth]{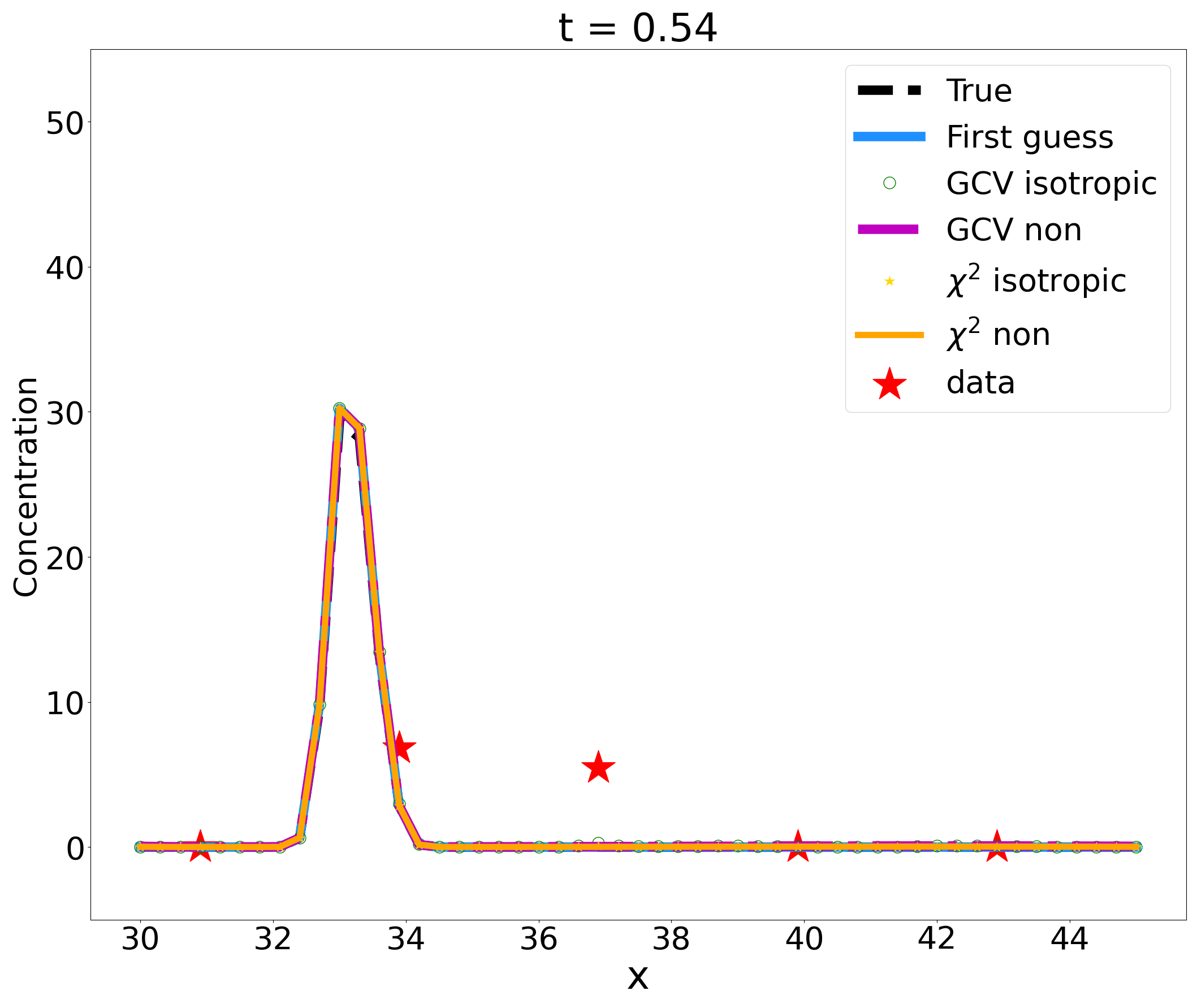}
    \end{subfigure}
    \begin{subfigure}[b]{0.49\textwidth}
        \includegraphics[width=\textwidth]{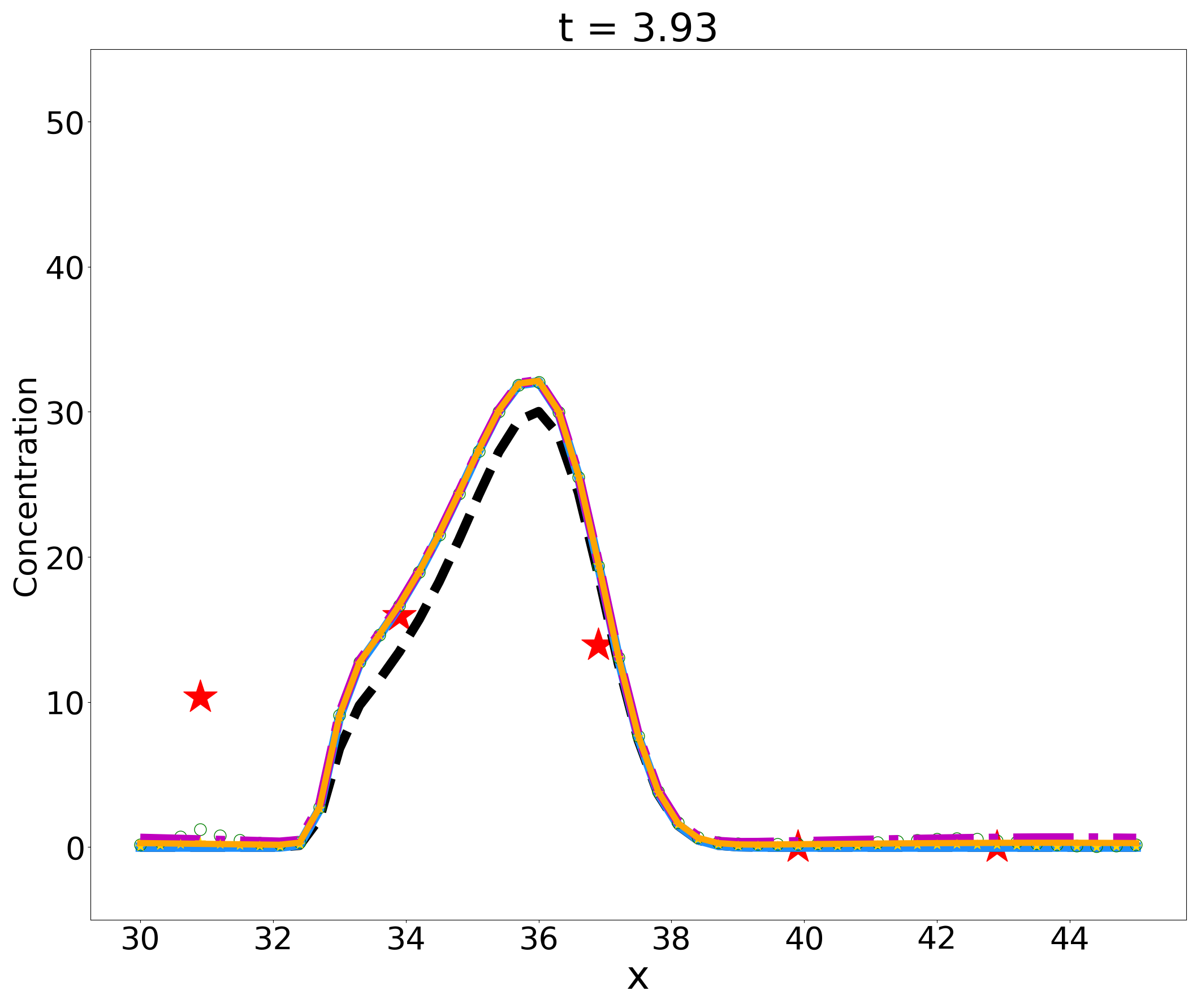}
    \end{subfigure}
    \begin{subfigure}[b]{0.49\textwidth}
        \includegraphics[width=\textwidth]{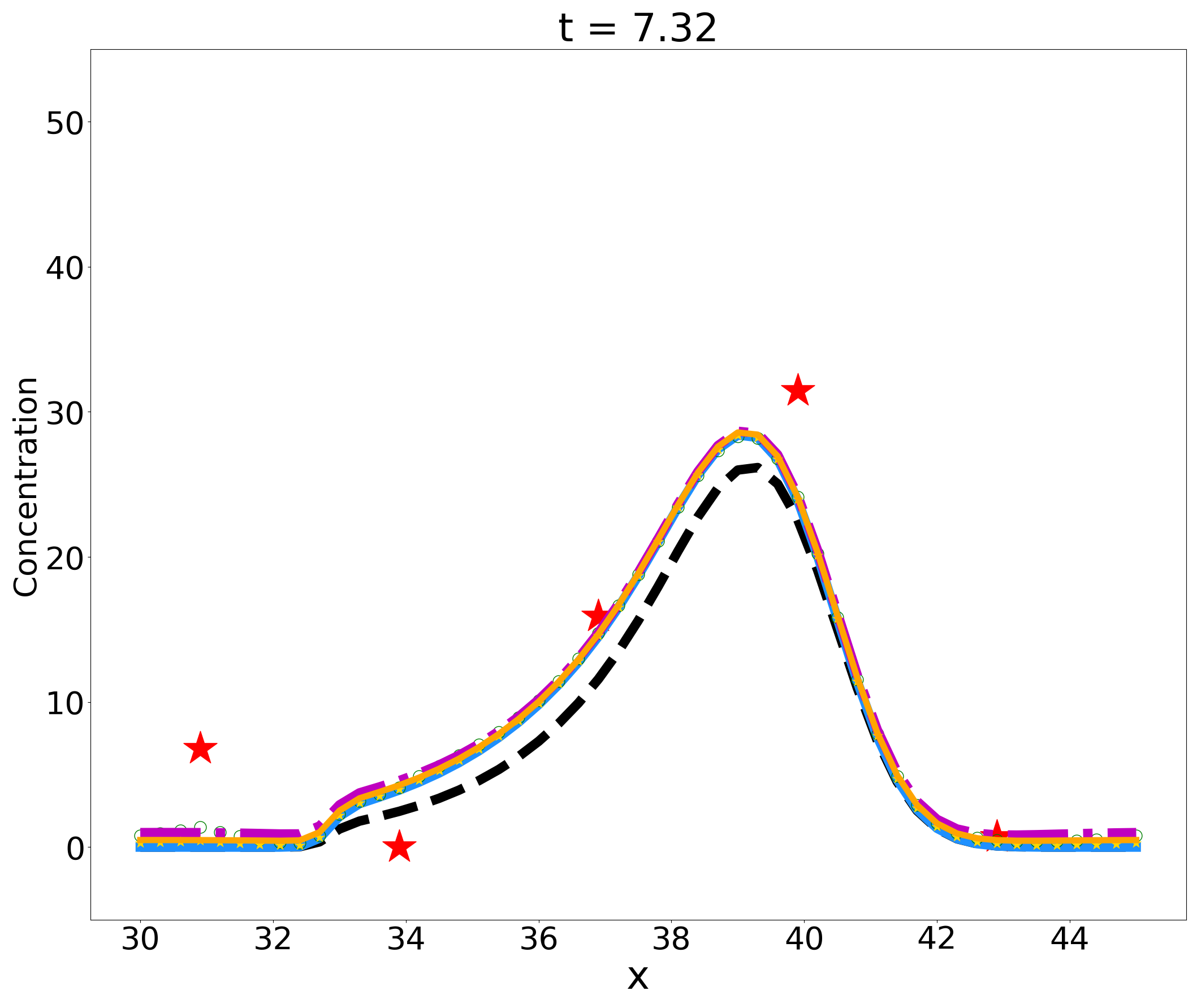}
    \end{subfigure}
    \begin{subfigure}[b]{0.49\textwidth}
        \includegraphics[width=\textwidth]{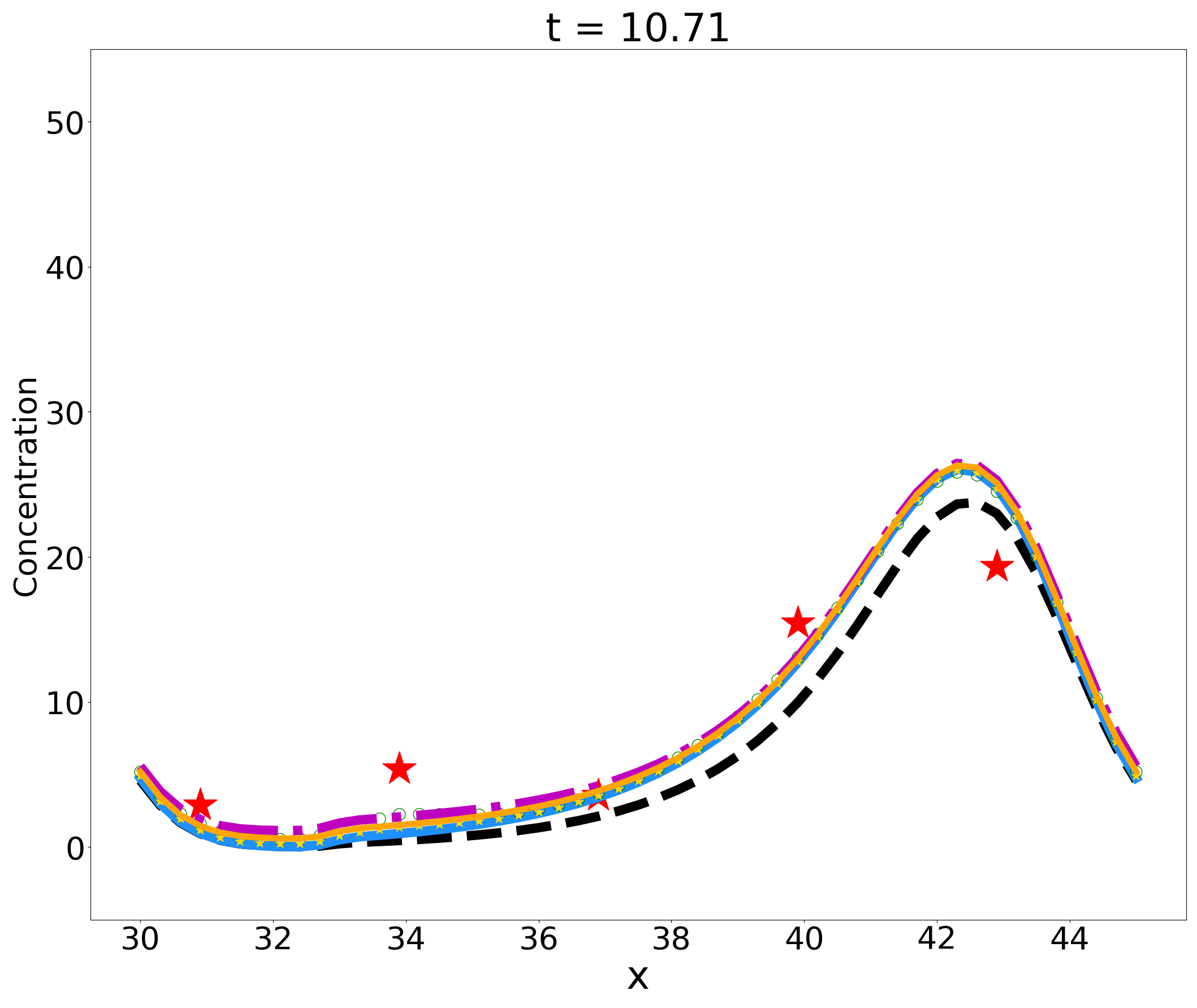}
    \end{subfigure}
    \caption{PM$_{2.5}$ concentration estimates as a function of space for \textit{experiment 1}}
   \label{fig:expt2_noniso}
\end{figure}

\textbf{Experiment 1.} \Cref{fig:expt2_noniso} shows the spatial PM$_{2.5}$ concentrations at four points in time during the transport process. The plots compare the true solution, first guess, observational data, and assimilated estimates obtained using the GCV and $\chi^2$ methods under both isotropic and non-isotropic covariance structures.

At $t=0.54$, all estimates, and the first guess, align closely with the true solution. Assimilated estimates from both isotropic and non-isotropic cases are nearly identical at this stage, reflecting minimal correction from the data. As the transport process evolves (at t=3.93), differences begin to emerge. The $\chi^2$ isotropic estimate remains almost identical to the first guess, while the other methods show small deviations. This behavior persists at later times  ($t = 7.32$ and $t = 10.71$), where the non-isotropic estimates offer slightly more responsiveness to the data but remain close to the first guess. Overall, the differences between the isotropic and non-isotropic methods are minimal in this case. All estimates remain close to the first guess, which is more accurate than the data, confirming that the estimated model error variances appropriately preserve model dominance in the assimilation. 
\FloatBarrier

\begin{figure}[H]
    \centering
    \begin{subfigure}[b]{0.49\textwidth}
        \includegraphics[width=\textwidth]{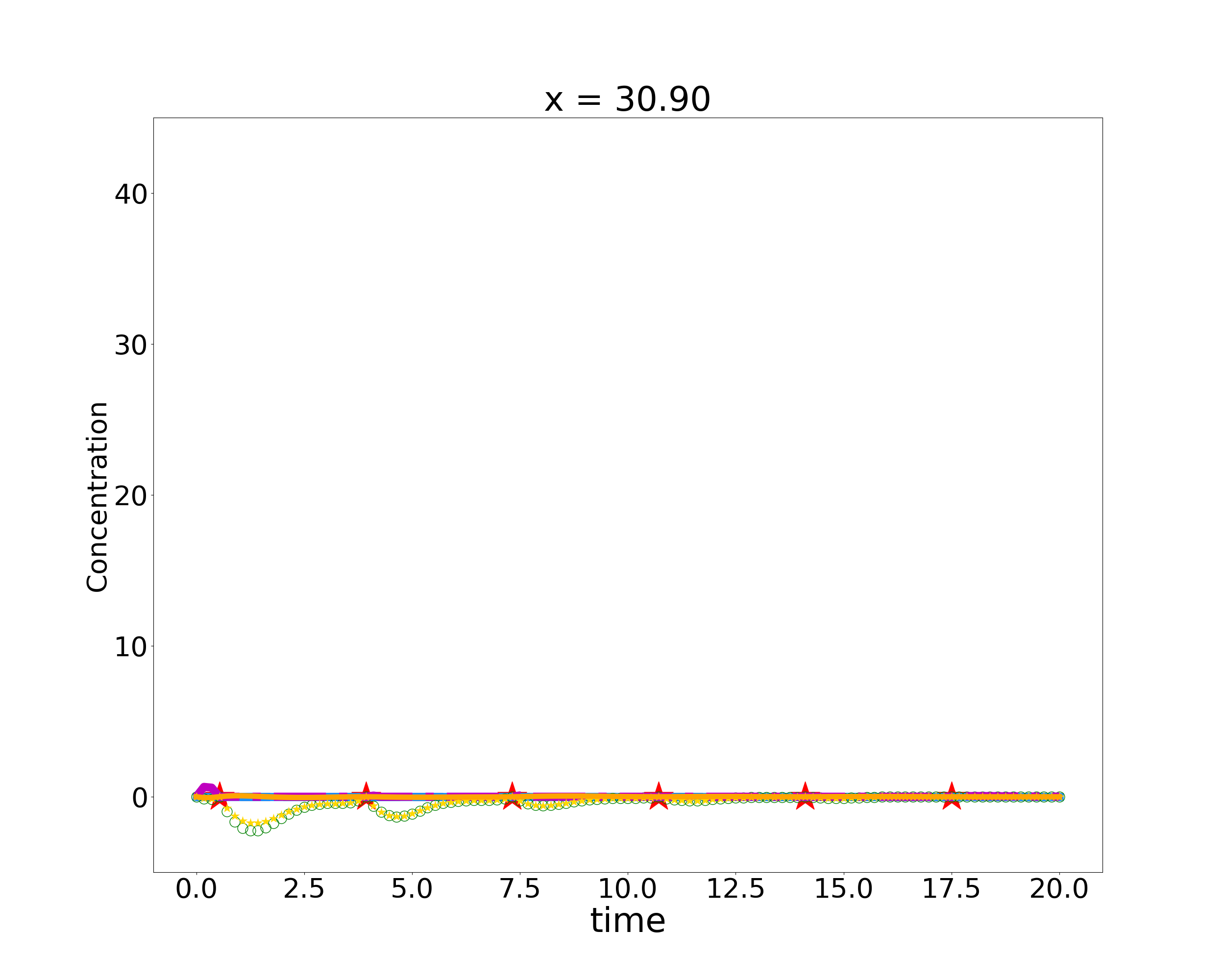}
    \end{subfigure}
    \begin{subfigure}[b]{0.49\textwidth}
        \includegraphics[width=\textwidth]{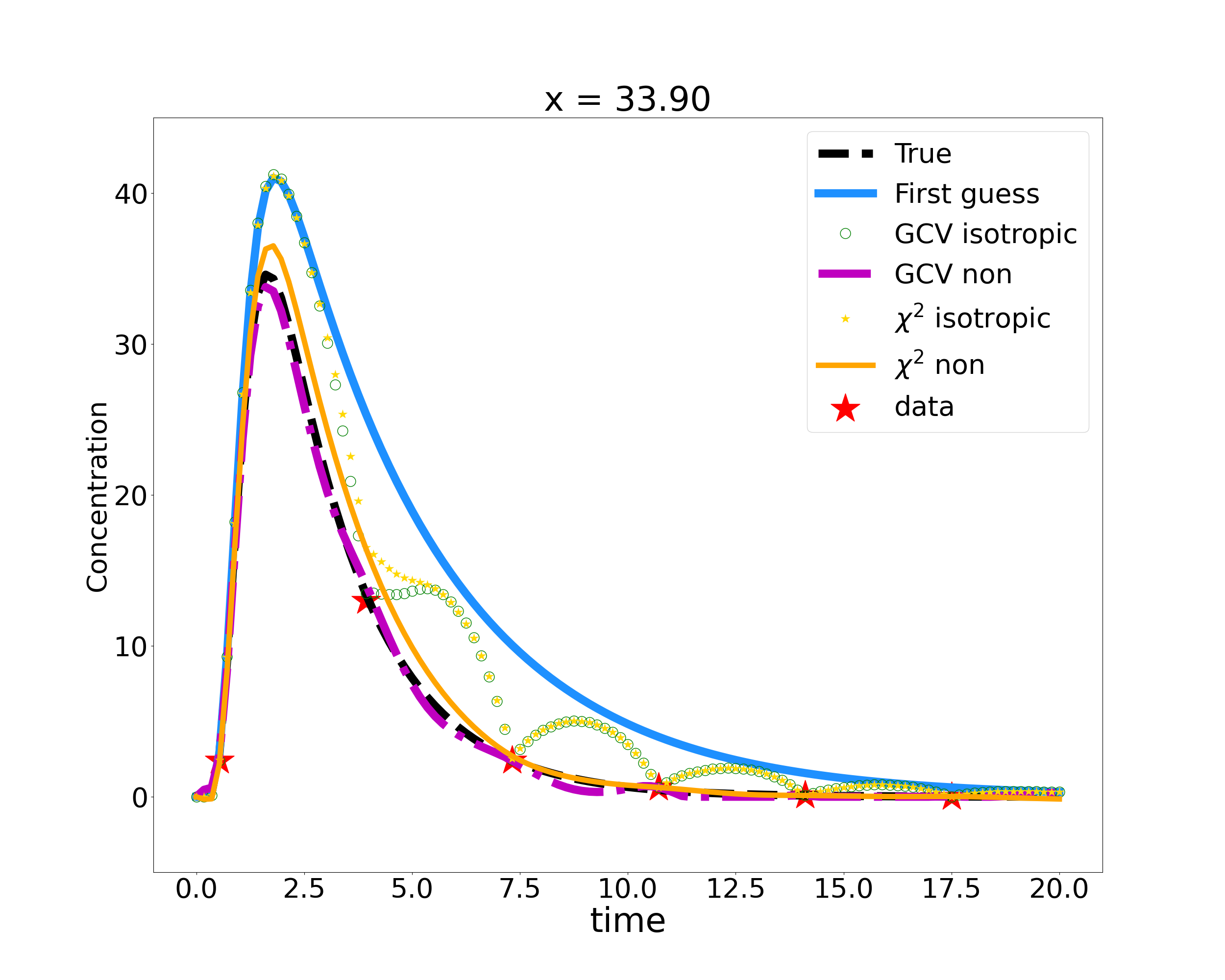}
    \end{subfigure}
    \begin{subfigure}[b]{0.49\textwidth}
        \includegraphics[width=\textwidth]{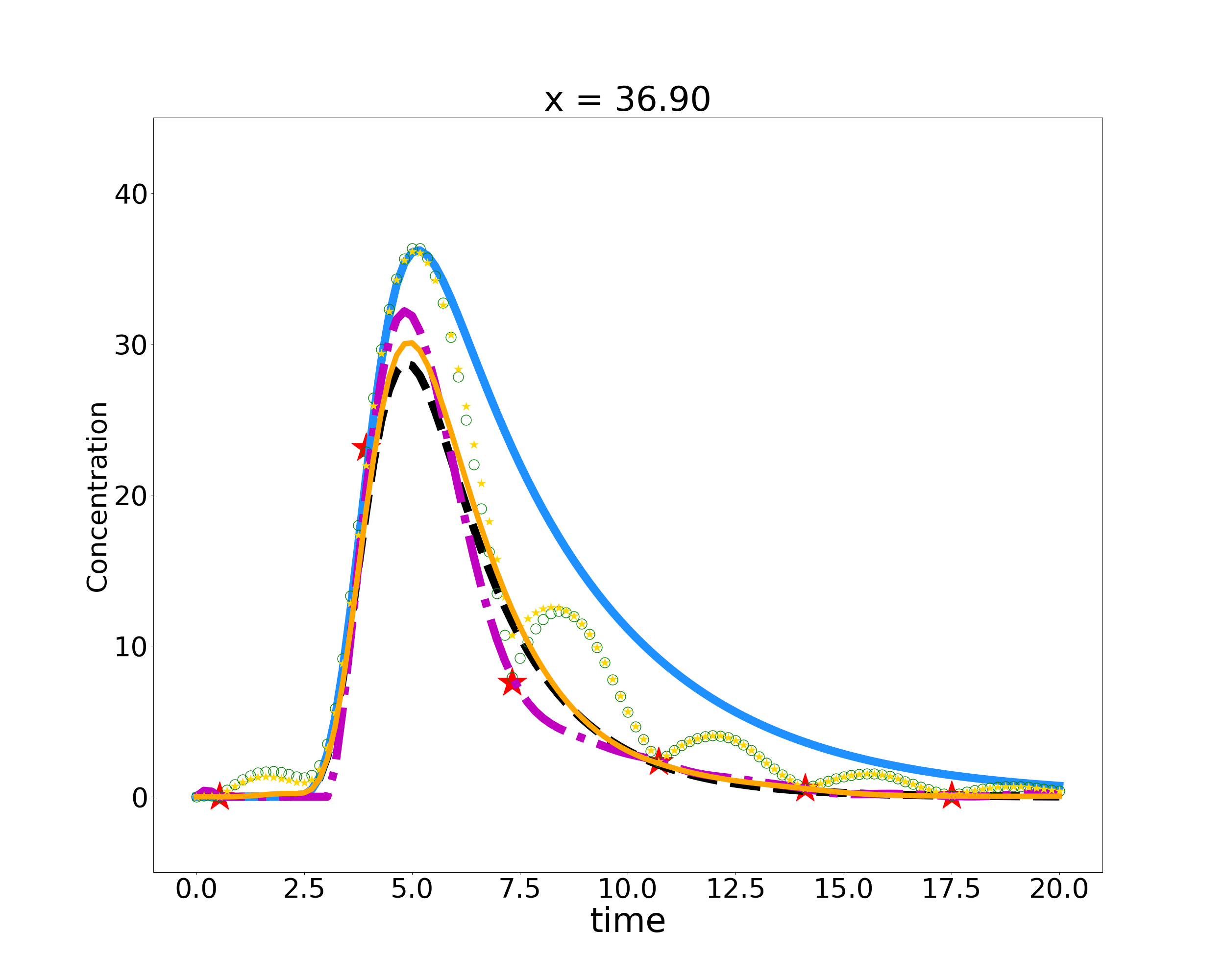}
    \end{subfigure}
    \begin{subfigure}[b]{0.49\textwidth}
        \includegraphics[width=\textwidth]{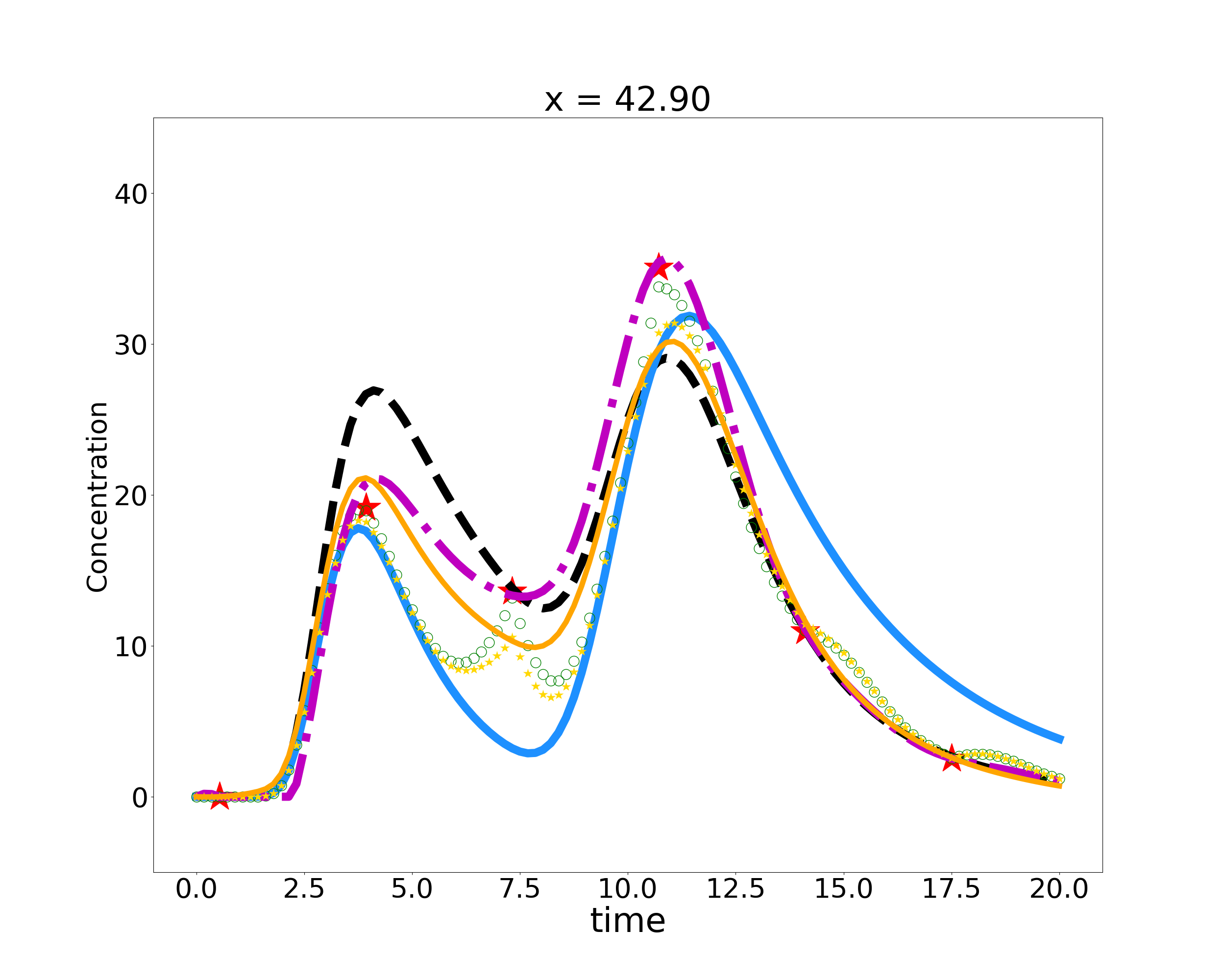}
    \end{subfigure}
    \caption{PM$_{2.5}$ concentration estimates as a function of time for \textit{experiment 4}}
    \label{fig:expt4_noniso}
\end{figure}

\textbf{Experiment 4.} \Cref{fig:expt4_noniso} presents the temporal evolution of PM$_{2.5}$ concentration estimates at different spatial locations. Also here, the plots compare the true solution, first guess, observational data, and assimilated estimates obtained using the GCV and $\chi^2$ methods under both isotropic and non-isotropic covariance structures.

At $x=30.90$, located upstream of the emission sources, the concentration remains near zero throughout the time interval, with only minor oscillations  by the assimilation estimates from the isotropic case. As the transport progresses, at $x=33.90$ and $x=36.90$, the peak begins to rise sharply and the differences between the methods become more evident. The non-isotropic estimates shift away from the first guess when needed and respond more accurately to the data, without introducing instability rather than the isotropic estimates which remain closer to the first guess at the peak. At $x=42.90$, where the plume is fully developed and the first guess deviates significantly from the true solution, the differences become more pronounced. The non-isotropic GCV and $\chi^2$ methods estimates successfully track the double-peak structure, fitting the data more closely and correcting both the timing and shape of the plume. The isotropic methods, on the other hand, produce smoother curves that miss key features and exhibit small oscillations due to lack of temporal coherence.

Overall, the non-isotropic methods consistently produce more accurate and stable estimates across all locations. They adapt more effectively to the data, correct deviations from the first guess where appropriate, and avoid the oscillations seen in the isotropic case. These improvements stem from the use of temporally correlated model error, which allows the non-isotropic approach to balance model and data information more intelligently throughout the assimilation process.

\FloatBarrier

\subsubsection{Root mean square errors}

\FloatBarrier
\begin{table}[tbhp]
\centering
\footnotesize
\renewcommand{\arraystretch}{1.2}
% \resizebox{\textwidth}{!}{%
\begin{tabular}{|c|c|c|c|c|c|c|} 
\hline
\textbf{Expt} & \textbf{First guess} & \textbf{Data} & \textbf{GCV isotropic} & \textbf{GCV non} & \boldmath$\chi^2$ \textbf{isotropic} & \boldmath$\chi^2$ \textbf{non} \\
\specialrule{1.2pt}{0pt}{0pt}
1 & 1.4673 & 4.2717 &  1.6689 & 1.9239 & 1.5248 & 1.7203 \\ \hline
2 & 2.1138 & 6.4341 & 2.1132 & 2.1071 &  2.1136 & 2.1112 \\ \hline
3 & 6.3103 & 3.1863 & 2.9821 & 2.4397 &  2.8786 & 2.5896 \\ \hline
4 & 5.7257 & 2.1447 &  3.5142 & 2.9429 & 3.6596 & 1.9554 \\
\specialrule{1.2pt}{0pt}{0pt}
\end{tabular}%
% }
\caption{RMSE between assimilated PM$_{2.5}$ estimates and the true concentration for each experiment.}
\label{tab:rmse_noniso}
\end{table}

\Cref{tab:rmse_noniso} presents the RMSE between the assimilated  PM$_{2.5}$ concentration estimates and the true solution for each experiment, comparing the first guess, data, and the GCV and $\chi^2$ methods under both isotropic and non-isotropic model error assumptions.

In experiments 1 and 2, where the first guess is more accurate than the data, the RMSEs from the non-isotropic methods are very similar to those from the isotropic case and closely match the RMSE of the first guess. The small differences indicate that the added flexibility of modeling spatial and temporal correlations offers limited benefit when the model already provides a strong first guess. In such cases, the isotropic covariance structure is sufficient to preserve model fidelity without overfitting the noisy data, making the non-isotropic formulation less impactful.

 In contrast, experiments 3 and 4, where the data are more accurate than the first guess, reveal a clear advantage for the non-isotropic error covariance. The RMSEs for both GCV and $\chi^2$ methods are substantially lower under the non-isotropic formulation. In experiment 3, the GCV multi method achieves the lowest RMSE (2.4397), significantly outperforming its isotropic counterpart (2.9821). A similar trend is observed in experiment 4, where the $\chi^2$ method yields the lowest RMSE across all methods (1.9554), outperforming both isotropic and the GCV non-isotropic estimates. These improvements highlight the non-isotropic framework's ability to more accurately capture spatial and temporal variations in the data, particularly when the first guess model is less reliable.

 Overall, the results suggest that when the model is already trusted more than the data such as in experiments 1 and 2, an isotropic covariance model is generally sufficient. However, in data-dominant scenarios with greater model uncertainty such as experiments 3 and 4, the non-isotropic framework offers clear benefits. By incorporating spatiotemporal correlations, it enables more accurate and physically realistic state estimates. These findings support the use of non-isotropic model error covariances in applications characterized by heterogeneous uncertainties or limited model accuracy.

\section{Conclusion}\label{sec:6}
In this work, we formulated weak constraint 4D-Var as a regularized inverse problem, with the model error covariance as the regularization matrix. The main focus of our study was on applying three regularization parameter selection methods namely, the L-curve, GCV, and $\chi^2$ methods to estimate hyperparameters for both isotropic and non-isotropic model error covariances.  A key aspect of our approach was the application of the representer method, which leverages the finite-dimensional nature of the observation space to express optimal state estimates as a linear combination of representers. This method reduces the solution search space from the state space to the data space, providing an efficient implementation for 4D-Var and enabling us to express the assimilated state estimates analytically. \textbf{Additionally, the representer method facilitated the derivation of matrix expressions for each regularization parameter selection method, such as $\mathcal{\hat{J}}_{\text{data}}$ in \cref{lemma1}, $\mathcal{\hat{J}}_{\text{mod}}$ in \cref{lemma2} for the L-curve, \cref{lemma1} for the GCV, and $\mathcal{\hat{J}}$ in \cref{theorem:3} for the $\chi^2$ method.}

We conducted numerical experiments using the transport equation to estimate PM${2.5}$ concentrations with simulated observational data. The results illustrated the importance of accounting for model imperfections in data assimilation. Across the four experiments, the  regularization parameter selection methods yielded consistent model error covariance estimates. These estimates accurately captured the balance between the influence of observational data and model predictions on assimilated state estimates. This balance is crucial for accurate PM$_{2.5}$ estimation in complex wildfire smoke transport scenarios, where, in addition to unknown uncertainties in the forcing, there are also unknown model physics.

In data-dominated regimes (Experiments 3 and 4), the estimated variances led to assimilated states that fit the data more closely, particularly when using the non-isotropic covariance structure. In contrast, when the model was more accurate (Experiments 1 and 2), the estimates preserved model fidelity and maintained proximity to the first guess. The non-isotropic framework provided improved flexibility and accuracy in data-dominated settings by incorporating spatial and temporal correlations in the model error. While the improvement was minimal when the model was already reliable, the non-isotropic structure became essential when the first guess was less trustworthy and the data carried more information. These results suggest that the isotropic approach may suffice in model-dominant settings, whereas the non-isotropic framework is preferred when observational data are more accurate or when uncertainty is heterogeneous.

In addition to providing optimal estimates of the covariance hyperparameters, the GCV and $\chi^2$ regularization parameter selection methods demonstrated practical efficiency in terms of computational cost. For the isotropic case, the GCV method typically required no more than 5 data assimilation runs, while the $\chi^2$ method converged within 7 runs. The L-curve method involves evaluating the residual and regularization norms over a predefined range of regularization parameters. In our implementation, we used a range of 100 values for the model error variance, which required 100 separate data assimilation runs. Due to this high computational cost, the L-curve method was not applied to the more intensive non-isotropic case. For the non-isotropic covariance case, GCV required up to 11 runs, and $\chi^2$ up to 29. These modest computational costs make the proposed framework viable for operational or real-time applications, even when extended to more complex systems.

Finally, although this work focused on estimating the model error covariance $C_f(\mathbf{x},t)$ under the assumption of perfect initial and boundary conditions, the same framework can be extended to jointly estimate the initial and boundary condition covariances $C_i (\mathbf{x})$ and $C_b(t)$ in future work. This paves the way toward fully data-driven error covariance estimation within weak constraint 4D-Var with representers. Future work will also focus on extending this approach to higher-resolution grids and larger ensemble sizes, leveraging the computational efficiency of pseudo-heat and boundary value problem-based methods in \cite{bennett1992inverse, egbert1994topex, ngodock2005efficient}. This approach will enable efficient application of spatiotemporal covariances without explicitly forming large covariance matrices.

%%%%%%%%%%%%%%%%%%%%%%%%%%%%%%%%%%%%%%%%%%%%%%%%%%%%%%%%%

\bibliographystyle{plain}
\bibliography{references}
% \bibliographystyle{siamplain}
% \bibliography{references}

\clearpage  % optional: ensures appendix starts on a new page
\input{appendix}

\end{document}

%% file: ex_shared.tex
\usepackage{lipsum}
\usepackage{amsfonts}
\usepackage{graphicx}
\usepackage{epstopdf}
\usepackage{algorithmic}
\usepackage{subcaption}
\ifpdf
  \DeclareGraphicsExtensions{.eps,.pdf,.png,.jpg}
\else
  \DeclareGraphicsExtensions{.eps}
\fi

\newsiamremark{remark}{Remark}
\newsiamremark{hypothesis}{Hypothesis}
\crefname{hypothesis}{Hypothesis}{Hypotheses}
\newsiamthm{claim}{Claim}

\headers{Model Error Covariance Estimation for Weak Constraint 4D-Var}{S. R. Babyale, J. Mead, D. Calhoun and P. O. Azike}

\title{Model Error Covariance Estimation for Weak Constraint Variational Data Assimilation\thanks{Submitted to the editors on September, 09 2024.
\funding{This work was funded by the National Science Foundation grant DMS-2111585.}}}

\author{SANDRA R. Babyale\thanks{Computing PhD, Boise State University, Boise, ID.} \and Jodi Mead\thanks{Department of  Mathematics, Boise State University, Boise, ID.}
\and Donna Calhoun\textsuperscript{\ddag} \and Patricia O. Azike\textsuperscript{\textdagger}}

\usepackage{amsopn}
\DeclareMathOperator{\diag}{diag}

\makeatletter
\newcommand*{\addFileDependency}[1]{% argument=file name and extension
  \typeout{(#1)}% latexmk will find this if $recorder=0 (however, in that case, it will ignore #1 if it is a .aux or .pdf file etc and it exists! if it doesn't exist, it will appear in the list of dependents regardless)
  \@addtofilelist{#1}% if you want it to appear in \listfiles, not really necessary and latexmk doesn't use this
  \IfFileExists{#1}{}{\typeout{No file #1.}}% latexmk will find this message if #1 doesn't exist (yet)
}
\makeatother

%% file: appendix.tex
\appendix
\section{Reduced Posterior Penalty Functionals}\label{appendix_A} In the weak-constraint 4D-Var framework with representers, the application of regularization parameter selection methods to estimate model error covariance hyperparameters requires explicit expressions for the penalty functionals evaluated at the optimal solution $\hat{q} (\mathbf{x},t)$. These functionals quantify the contributions of model dynamics and data misfit, forming the basis for selecting an optimal regularization hyperparameter. Using the analytical expression for $\hat{q} (\mathbf{x},t)$ in (\ref{3.15}), we derive the reduced penalty functionals, which enable the application of the L-curve, GCV, and $\chi ^2$ method. The results in this section are given in \cite{bennett1992inverse}. For simplicity and generalizability, we assume isotropic model and initial condition errors and omit the boundary condition term.

Let the penalty functional for the model be defined as 
\begin{align}
    \begin{aligned}
        \mathcal{\hat{J}}_{\text{mod}}\equiv \mathcal{{J}}_{\text{mod}} [\hat{q}] = {}& W_f \int_{0}^{T}\int_{\Omega}\bigg\{\frac{\partial \hat{q}}{\partial t} + L  \hat{q}(\mathbf{x},t) - Q(\mathbf{x},t)\bigg\}^2 d\mathbf{x} dt \\&
        + W_i \int_{\Omega}\big\{\hat{q}(\mathbf{x},0) - I(\mathbf{x})\big\}^2 d\mathbf{x} 
    \end{aligned} \label{C311_1}
\end{align}
 and that for the data as
 \begin{equation}
     \mathcal{\hat{J}}_{\text{data}} \equiv \mathcal{{J}}_{\text{data}} [\hat{q}] = \sum_{m=1}^{M} w_m \big\{ \hat{q}(\mathbf{x}_m,t_m) - d_m \big\}^2 \label{C311_2}
 \end{equation}
so that $$\mathcal{\hat{J}} \equiv\mathcal{{J}} [\hat{q}] = \mathcal{\hat{J}}_{\text{mod}} + \mathcal{\hat{J}}_{\text{data}},$$
where $W_f$ and $W_i$ are the inverse variances of the model and initial condition errors, respectively. 

\begin{lemma} \label{lemma1}
 The data penalty satisfies 
    \begin{equation}    \mathcal{\hat{J}}_{\text{data}}  = \mathbf{h}^T\mathbf{P}^{-1}\mathbf{W}_{\text{d}}^{-1}\mathbf{P}^{-1}\mathbf{h} \label{18}
    \end{equation}
\end{lemma}
\begin{proof}
    Writing (\ref{C311_2}) in matrix form we have
        \begin{equation}  \mathcal{\hat{J}}_{\textnormal{data}}  = \big(\mathbf{\hat{q}} - \mathbf{d} \big)^T\mathbf{W}_d\big(\mathbf{\hat{q}} - \mathbf{d} \big)\label{20}.
    \end{equation}
    Subsititing equation (\ref{3.11}) into the Euler-Lagrange equations (\ref{3.5}) and equating coefficients of the impulses, the representer coefficients $\boldsymbol{\hat{\beta}}$ can be written as $\boldsymbol{\hat{\beta}} = \mathbf{W}_d(\mathbf{\hat{q}} - \mathbf{d})$. Thus 
    \begin{equation}  
        \mathcal{\hat{J}}_{\text{data}}   = \boldsymbol{\hat{\beta}} ^T \mathbf{W}_d^{-1} \boldsymbol{\hat{\beta}}.\label{21}
    \end{equation}
    Using (\ref{10}), we have 
    \begin{equation}  
        \mathcal{\hat{J}}_{\text{data}}    = \big((\mathbf{R}+\mathbf{W}_d^{-1})^{-1}\mathbf{h} \big)^T\mathbf{W}_d^{-1}\big((\mathbf{R}+\mathbf{W}_d^{-1})^{-1}\mathbf{h}\big)
     \label{22}
    \end{equation}
    Noting that $\mathbf{P} = \mathbf{R} + \mathbf{W} _d ^{-1}$ is  symmetric, the result then follows.  
\end{proof}
\begin{lemma} \label{lemma2}
    The model penalty satisfies
\begin{equation}
\mathcal{\hat{J}}_{\text{mod}} =  \mathbf{h}^T\mathbf{P} ^{-1} \mathbf{R} \mathbf{P} ^{-1} \mathbf{h} \label{25}
\end{equation}
\end{lemma}
\begin{proof}
    Substituting for $\hat{q}(\mathbf{x},t)$ using (\ref{3.15}) into the penalty functional (\ref{C311_1}), we have 
\begin{align}
    \begin{aligned}
        \mathcal{\hat{J}}_{\text{mod}}= {}& W_f \int_{0}^{T} \int_{\Omega} \bigg\{
        \frac{\partial q_F}{\partial t} + L q_F(\mathbf{x},t)  - Q(\mathbf{x},t)  \\&
        \quad +  \mathbf{h}^T \mathbf{P}^{-1} \left(\frac{\partial \mathbf{r}(\mathbf{x},t)}{\partial t} + L \mathbf{r}(\mathbf{x},t)\right)
        \bigg\}^2 d\mathbf{x} dt \\& 
        \quad\quad + W_i \int_{\Omega} \big\{q_F(\mathbf{x},0)- I(\mathbf{x}) + \mathbf{h}^T \mathbf{P}^{-1} \mathbf{r}(\mathbf{x},0)\big\}^2 d\mathbf{x}
    \end{aligned}
    \label{26}
\end{align}

Since the first guess $q_F(\mathbf{x},t)$ is the solution to the error-free model (\ref{3.1}), we have
\begin{align}
    \begin{aligned}
        \mathcal{\hat{J}}_{\text{mod}} 
        ={}& \mathbf{h}^T\mathbf{P} ^{-1} \bigg \{  W_f\int_{0}^{T} \int_{\Omega} \bigg(\frac{\partial \mathbf{r}}{\partial t} + L \mathbf{r}(\mathbf{x},t)\bigg)\bigg(\frac{\partial \mathbf{r}}{\partial t} + L \mathbf{r}(\mathbf{x},t)\bigg)^T d\mathbf{x} dt \\& + W_i 
         \int_{\Omega}  \mathbf{r}(\mathbf{x},0)\mathbf{r}(\mathbf{x},0)^T d\mathbf{x}  + W_b 
         \int_{0}^{T}  \mathbf{r}(\mathbf 0,t)\mathbf{r}(\mathbf 0,t)^T dt\bigg \} \mathbf {P} ^{-1}\mathbf{h}
    \end{aligned}\label{27}
\end{align}
Given that $\mathbf{r}(\mathbf{x},t)$  is the solution to ($\ref{8}$),
\begin{align}
    \begin{aligned}
        \mathcal{\hat{J}}_{\text{mod}} 
         &= \mathbf{h}^T\mathbf{P} ^{-1} \bigg \{  W_f^{-1}\int_{0}^{T} \int_{\Omega} \boldsymbol{\alpha}(\mathbf{x},t)\boldsymbol{\alpha}(\mathbf{x},t)^T d\mathbf{x} dt  \\&+ 
         W_i^{-1} \int_{\Omega}  \boldsymbol{\alpha}(\mathbf{x},0)\boldsymbol{\alpha}(\mathbf{x},0)^T d\mathbf{x}  \bigg \} \mathbf {P} ^{-1}\mathbf{h}
    \end{aligned}\label{28}
\end{align}
where $\boldsymbol{\alpha} (\mathbf{x},t)= [{\alpha}_1(\mathbf x, t), {\alpha}_2(\mathbf x, t), \cdots {\alpha}_M(\mathbf x, t)]^T$.
The adjoint representer function $\alpha_m(\mathbf{x},t)$ for a point measurement $(\mathbf{x}_m,t_m)$ is the Green's function $\gamma (\mathbf{x},t,\mathbf{x}_m,t_m)$ where $\gamma (\mathbf{x},t,\mathbf{y},s)$ satsifies 
\begin{align}
\begin{aligned}
\mathcal{L}^*\gamma (\mathbf{x},t,\mathbf{y},s) \equiv -\frac{\partial \gamma (\mathbf{x},t,\mathbf{y},s)}{\partial t} - L^T \gamma (\mathbf{x},t,\mathbf{y},s) &= \delta(\mathbf{x} - \mathbf{y})\delta(t-s)\\
\gamma (\mathbf{x},T,\mathbf{y},s) &=0
\label{29}
\end{aligned}
\end{align}
The corresponding representer functions  $r_m(\mathbf x,t) = \Gamma (\mathbf{x},t,\mathbf{x_m},t_m) $ where
\begin{align}
\begin{aligned}
\mathcal{L} \Gamma (\mathbf{x},t,\mathbf{y},s) \equiv \frac{\partial \Gamma (\mathbf{x},t,\mathbf{y},s)}{\partial t} + L\Gamma (\mathbf{x},t,\mathbf{y},s)&= W_f^{-1} \gamma (\mathbf{x},t,\mathbf{y},s)\\
\Gamma (\mathbf{x},0,\mathbf{y},s) &= W_i^{-1} \gamma (\mathbf{x},0,\mathbf{y},s)
\label{30}
\end{aligned}
\end{align}
Integrating by parts and using dummy variables $\mathbf z$ and $r$, we obtain 
\begin{align}
  \begin{aligned}
    \int_{0}^{T} \int_{\Omega} \gamma (\mathbf{x},t,\mathbf{z},r)\mathcal{L}\Gamma (\mathbf{z},r,\mathbf{y},s) \,d\mathbf{z}\,dr = {}& \int_{\Omega} \gamma (\mathbf{x},t,\mathbf{z},r) \Gamma (\mathbf{z},r,\mathbf{y},s) \bigg|_0^T \,d\mathbf{z} \\&
    \quad + \int_{0}^{T} \int_{\Omega} \Gamma (\mathbf{x},t,\mathbf{z},r)\mathcal{L}^*\gamma (\mathbf{z},r,\mathbf{y},s) \,d\mathbf{z}\,dr
  \end{aligned}
\end{align}
Substituting $\mathcal{L}^*\gamma (\mathbf{z},r,\mathbf{y},s) = \delta(\mathbf{z}-\mathbf{y})\delta(r-s)$ we obtain
\begin{align}
  \begin{aligned}
    \Gamma (\mathbf{x},t,\mathbf{y},s) ={}& \int_{0}^{T} \int_{\Omega} \gamma (\mathbf{x},t,\mathbf{z},r) \mathcal{L}\Gamma (\mathbf{z},r,\mathbf{y},s)\,d\mathbf{z}\,dr  - \int_{\Omega} \gamma (\mathbf{x},t,\mathbf{z},r) \Gamma (\mathbf{z},r,\mathbf{y},s) \bigg|_0^T \,d\mathbf{z} 
  \end{aligned}
\end{align}
Substituting for $\mathcal{L}\Gamma (\mathbf{z},r,\mathbf{y},s)$ and $\Gamma (\mathbf{z},0,\mathbf{y},s)$ and  evaluating at the data points, we have
\begin{align}
    \begin{aligned}
        \Gamma (\mathbf{x}_m,t_m,\mathbf{x}_l,t_l,) = {}& W_f^{-1}\int_{0}^{T} \int_{\Omega} \gamma (\mathbf{x}_m,t_m,\mathbf{z}_m,r_m)\gamma (\mathbf{z}_m,r_m,\mathbf{x}_l,t_l,) d\mathbf{z} dr \\& + 
         W_i^{-1} \int_{\Omega}  \gamma (\mathbf{x}_m,t_m,\mathbf{z}_m,0)\gamma (\mathbf{z}_m,0,\mathbf{x}_l,t_l,) d\mathbf{z} 
    \end{aligned}
\end{align}
\begin{align}
    \begin{aligned}
        \Gamma (\mathbf{x}_m,t_m,\mathbf{x}_l,t_l) = {}& W_f^{-1}\int_{0}^{T} \int_{\Omega} \alpha_l(\mathbf{x}_m,t_m)\alpha_l(\mathbf{x}_m,t_m) d\mathbf{x} dt \\&  + W_i^{-1} \int_{\Omega}  \alpha_l(\mathbf{x}_m,0)\alpha_l(\mathbf{x}_m,0) d\mathbf{x} \\
        ={}& \mathbf{R}_{lm}
    \end{aligned}
\end{align}
If we consider for all $l,m$ we have 
\begin{align}
    \begin{aligned}
        \mathbf{R} = {}& W_f^{-1}\int_{0}^{T} \int_{\Omega} \boldsymbol{\alpha}(\mathbf{x},t)\boldsymbol{\alpha}(\mathbf{x},t)^T d\mathbf{x} dt  + 
         W_i^{-1} \int_{\Omega}  \boldsymbol{\alpha}(\mathbf{x},0)\boldsymbol{\alpha}(\mathbf{x},0)^T d\mathbf{x}  
    \end{aligned}
\end{align}
\end{proof}

\begin{theorem} \label{theorem:3}
   The posterior functional $\mathcal{\hat{J}} $ satisfies
\end{theorem}
\begin{equation}
\hat{\mathcal{J}}  
=  \mathbf{h}^T\mathbf{P}^{-1}\mathbf{h} \label{16}
\end{equation}

\begin{proof}
Using lemmas (\ref{lemma1}) and (\ref{lemma2}) we have 
\begin{align}
    \begin{aligned}
        \mathcal{\hat{J}} & =  \mathbf{h}^T \mathbf{P}^{-1} \mathbf{W}_d^{-1} \mathbf{P}^{-1} \mathbf{h} + \mathbf{h}^T\mathbf{P}^{-1} \mathbf{R} \mathbf{P}^{-1} \mathbf{h}\\
        & = \mathbf{h}^T\mathbf{P}^{-1} \big(\mathbf{R} + \mathbf{W}_d^{-1}\big)\mathbf{P} ^{-1} \mathbf{h}\\
        & = \mathbf{h}^T\mathbf{P} ^{-1} \mathbf{h}.
    \end{aligned}
\end{align}
\end{proof}

\section{Proof of \cref{thm:4}}\label{appendix_C}
\begin{proof}
Consider a modified weak constraint 4D-Var data assimilation problem where we leave out a data point $d_k$:
\begin{multline} \label{C3_9}
     \min_ {q}\bigg\{ \sum_{m\neq k}^M{w_m}\big[ q(\mathbf x_m,t_m) - d_m \big]^2  + C_{i}^{-1}\int_{\Omega}{\big[q(\mathbf{x},0) - I(\mathbf{x})\big]^2} d\mathbf{x}  \\ \quad + C_{b}^{-1}\int_{0}^{T}\big[q(\mathbf{0},t)-B(t)\big]^2 dt + {C_f}^{-1}\int_{0}^{T}\int_{\Omega} \left[\frac{\partial {q}}{\partial t} + Lq(\mathbf{x},t) - Q\right]^2 d\mathbf x dt \bigg \},
      \end{multline}
where $w_m = \sigma ^{-2}_m$. Let $\hat{q}^{[k]}(\mathbf x,t)$ be the solution to the optimization problem (\ref{C3_9}), where the superscript $k$ indicates that $d_k$ was left out of the computation. In the leave-one-out approach, we select the regularization parameter $C_f = \sigma_f^{2}$ that minimizes the sum of predictive errors over all $k$:
\begin{align}
        \begin{aligned}
     \min_{\sigma _f} g(\sigma _f) \equiv \min_{\sigma _f} \bigg \{\frac{1}{M} \sum_{k = 1}^M {w_k} \big(\hat{q}^{[k]}(\mathbf x_k,t_k) - d_k\big)^2 \bigg\}.
        \end{aligned} \label{C3_10}
      \end{align}
Computing $g(\sigma_f)$ involves solving $M$ problems of the form (\ref{C3_9}) and as described in \cite{golub1979generalized}, we can speed up this computation by using the leave-one-out lemma. In the case of weak constraint 4D-Var, note that $\hat{q}^{[k]}(\mathbf x,t)$ optimizes 
\begin{multline}  \label{C3_12}
    \min\limits_q \bigg\{ w_k \big(q(\mathbf x_k, t_k) - \tilde{d}_k\big)^2 + \sum\limits_{m \neq k} w_m \big(q(\mathbf x_m, t_m) - \tilde{d}_m\big)^2  + C_{i} ^{-1}\int_{\Omega}{\big[q(\mathbf{x},0) - I(\mathbf{x})\big]^2} d\mathbf{x}  \\ \quad + C_{b}^{-1} \int_{0}^{T}\big[q(\mathbf{0},t)-B(t)\big]^2 dt + C_f ^{-1} \int_{T_0}^T \int_\Omega \bigg[\frac{\partial q}{\partial t} + Lq(\mathbf x,t) - Q(\mathbf x,t)\bigg]^2 \,d\mathbf x\,dt \bigg\}
\end{multline}
where \begin{equation}
\tilde{d}_m = \begin{cases} \hat{q}^{[k]}(\mathbf x_k, t_k) & m = k \\ d_m & m \neq k\end{cases}. \end{equation}
Define $\hat{\mathbf{q}}_m = \begin{bmatrix}
    \hat{q}(\mathbf x_1,t_1) & \hat{q}(\mathbf x_2,t_2)& \cdots & \hat{q}(\mathbf x_M,t_M)
\end{bmatrix} ^T$ then $$
     \hat{\mathbf{q}}_m = {\mathbf{q}_F}_m + \mathbf{R} \mathbf{P}^{-1}(\mathbf{{\mathbf d}} - {\mathbf{q}_F}_m)~ \text{and} ~ \hat{\mathbf{q}}_m^{[k]} = {\mathbf{q}_F}_m + \mathbf{R} \mathbf{P}^{-1}(\mathbf{\tilde{\mathbf d}} - {\mathbf{q}_F}_m).$$
Consider $\hat{\mathbf{q}}_m^{[k]} - \hat{\mathbf{q}}_m = \mathbf{R} \mathbf{P}^{-1}(\mathbf{\tilde{\mathbf d}} - \mathbf d)$ so that
\begin{equation}
     \hat{{q}}^{[k]}(\mathbf{x}_k,t_k) - \hat{{q}}(\mathbf{x}_k,t_k) = (\mathbf{R} \mathbf{P}^{-1}\tilde{\mathbf d})_k - (\mathbf{R} \mathbf{P}^{-1}{\mathbf d})_k.\label{C3_14_2}
\end{equation}
Since $\tilde{d}_m = d_m$ where $m\neq k$ and 
\begin{equation}
    (\mathbf{R} \mathbf{P}^{-1}\tilde{\mathbf{d}})_k = \sum_{\substack{m = 1 \\ m \neq k}}^M (\mathbf{R} \mathbf{P}^{-1})_{km}d_k + (\mathbf{R} \mathbf{P}^{-1})_{kk} \tilde{d}_k,
\end{equation}
we have that
\begin{equation}
     \hat{{q}}^{[k]}(\mathbf{x}_k,t_k) - \hat{{q}}(\mathbf{x}_k,t_k) = (\mathbf{R} \mathbf{P}^{-1})_{kk} (\tilde{ d}_k - {d}_k).
\end{equation}
Now 
\begin{align}
    \begin{aligned}
    \frac{{\Tilde{ d}}_k - {d}_k - \hat{q}^{[k]}(\mathbf x_k,t_k) +  \hat{q} (\mathbf x_k,t_k)}{{\Tilde{ d}}_k - {d}_k} = 1 - (\mathbf{R} \mathbf{P}^{-1})_{kk}
     \label{C3_19}
    \end{aligned}
\end{align} 
and since $\hat{q}^{[k]}(\mathbf x_k,t_k) = \Tilde{ d}_k$, we have
\begin{align}
    \begin{aligned}
    \frac{\hat{q} (\mathbf x_k,t_k) - {d}_k }{\hat{q}^{[k]}(\mathbf x_k,t_k) - {d}_k} &= 1 - (\mathbf{R} \mathbf{P}^{-1})_{kk}.
    \end{aligned}
\end{align} 
This means we do not have to solve the data assimilation problem $k$ times because
\begin{align}
    \begin{aligned}
    \hat{q}^{[k]}(\mathbf x_k,t_k) - {d}_k &= \frac{\hat{q} (\mathbf x_k,t_k) - {d}_k}{1-(\mathbf{R} \mathbf{P}^{-1})_{kk}}.
     \label{C3_20}
    \end{aligned}
\end{align} 
Substituting for $\hat{q}^{[k]}(\mathbf x_k,t_k) - {d}_k$ in (\ref{C3_10}), we find the optimal $\sigma _f ^2$ by minimizing
\begin{align}
    \begin{aligned}
          g(\sigma _f) = \frac{1}{M} \sum_{k = 1}^M w_k\left(\frac{\hat{q} (\mathbf x_k,t_k) - {d}_k}{1-(\mathbf{R} \mathbf{P}^{-1})_{kk}}\right)^2. 
    \end{aligned} \label{gcv_12}
\end{align}
% Simplifying further, we replace $(\mathbf{R} \mathbf{P}^{-1})_{kk}$ with it's average value and we minimize
% \begin{align}
%     g(\sigma _f) &= M  \frac{(\hat{\mathbf q}_m- \mathbf d)^T\mathbf{C}_{\epsilon}^{-1} (\hat{\mathbf q}_m- \mathbf d)}{\text{Tr}(\mathbf{I}-\mathbf{R}\mathbf{ P}^{-1})^2}\\
%     &=   \frac{M\mathcal{\hat{J}}_{\text{data}}}{\text{Tr}(\mathbf{I}-\mathbf{R}\mathbf{ P}^{-1})^2}.\label{gcv_1}
% \end{align}
% Applying \cref{lemma1}, the result follows. 
The optimal model error variance $\sigma_f$ is the one which minimize (\ref{gcv_12}).
\end{proof}

\section{Upwind finite volume method}\label{appendix_B}
We solve the transport model using the upwind finite volume method. To solve the transport equation (\ref{C4_1}) using this method, the spatial domain ${x} \in [30,45]$ is divided into control volumes (grid cells) and each control volume has a width $\Delta x$ \cite{leveque2002finite}. The time interval $t = [0,20]$ is discretized into time steps $t_n$, where $\Delta t = t_{n+1}-t_n$. For the $i^{th}$ control volume with a center $x_i$, the cell-averaged value of $q(x,t)$ at time $t_n$ is denoted by $q_i^n$ which approximates 
\begin{equation}
   q_i^n \approx  \frac{1}{\Delta x} \int_{x_{i-\frac{1}{2}}}^{x_{i+\frac{1}{2}}} q(x,t_n)dx.
\end{equation}
The finite volume method integrates the transport equation over each control volume $\big[x_{i+\frac{1}{2}}, x_{i-\frac{1}{2}}\big]$, giving:
\begin{equation}
    \frac{d}{dt} \int_{x_{i-\frac{1}{2}}}^{x_{i+\frac{1}{2}}} q(x,t)dx = - \int_{x_{i-\frac{1}{2}}}^{x_{i+\frac{1}{2}}} u\frac{dq}{dx}dx + \int_{x_{i-\frac{1}{2}}}^{x_{i+\frac{1}{2}}} Q(x,t)dx. \label{up_1}
\end{equation}
Using the forward Euler method for the time derivative and approximating the fluxes at the cell interfaces with the upwind scheme, the fully discrete form of the transport equation becomes:
\begin{equation}
    q_i^{n+1} = q_i^n - \frac{\Delta t}{\Delta x} \left( F_{i+\frac{1}{2}}^n - F_{i-\frac{1}{2}}^n \right) + \Delta t Q_i^n,
\end{equation}
where  \( F_{i+\frac{1}{2}}^n \) is the flux at the right interface of cell \( i \), \( F_{i-\frac{1}{2}}^n \) is the flux at the left interface of cell \( i \) and \( Q_i^n \) is the source term averaged over the cell at time \( t_n \).

This scheme accounts for the direction of the wind field $u$ which is key for solving the forward (\ref{8}) and backward (\ref{9}) systems that result from using the method of representers. If the wind field $u >0$, the advection term at the interface $x_{i+\frac{1}{2}}$ is influenced by the upstream value (left side of the interface). Conversely, if $u<0$ , it is influenced by the downstream value (right side of the interface). Thus, the upwind flux \( F_{i+\frac{1}{2}}^n \) depends on the sign of velocity $u$:

\begin{equation}
    F_{i+\frac{1}{2}} = \begin{cases}
        u q_i,~~~~~~ \text{if}~~ u>0\\
        u q_{i+1},~~~~~~ \text{if}~~ u<0.
    \end{cases}
\end{equation}
The time step $\Delta t$ is chosen so that the Courant-Friedrichs-Lewy (CFL) condition is satisified.